\documentclass[a4paper,12pt]{article}
\usepackage[latin1]{inputenc}
\usepackage{mathtools}
\usepackage{amsfonts,amssymb,amsmath,dsfont} 
\usepackage{natbib}
\usepackage{graphicx}
\usepackage{array} 
\usepackage[nottoc]{tocbibind} 
\usepackage{hyperref} 
\usepackage{color}
\usepackage{authblk}
\usepackage{fullpage}
\renewcommand*{\thefootnote}{\fnsymbol{footnote}}
\usepackage[font=footnotesize]{caption,subcaption} 

\begin{document}
\title{\vspace{-30pt}Residential income segregation:\protect\\ A behavioral model of the housing market }
\author[1,2]{Marco Pangallo$^{\star,}$}
\author[3,4]{Jean-Pierre Nadal}
\author[3,5]{Annick Vignes}
\affil[1]{\small Institute for New Economic Thinking at the Oxford Martin School, University of Oxford, Oxford OX26ED, UK}
\affil[2]{Mathematical Institute, University of Oxford, Oxford OX13LP, UK}
\affil[3]{Centre d'Analyse et de Math\'ematique Sociales, \'Ecole des Hautes \'Etudes en Sciences Sociales, 54 bd Raspail, 75006 Paris, France}
\affil[4]{Laboratoire de Physique Statistique, \'Ecole Normale Sup\'erieure, 24 rue Lhomond, 75231 Paris, France}
\affil[5]{\'Ecole des Ponts ParisTech, 6-8 Avenue Blaise Pascal, 77455 Champs-sur-Marne, France}

\date{\today}
\maketitle
\abstract{We represent the functioning of the housing market and study the relation between income segregation, income inequality and house prices by introducing a spatial Agent-Based Model (ABM). Differently from traditional models in urban economics, we explicitly specify the behavior of buyers and sellers and the price formation mechanism. Buyers who differ by income select among heterogeneous neighborhoods using a probabilistic model of residential choice; sellers employ an aspiration level heuristic to set their reservation offer price; prices are determined through a continuous double auction. We first provide an approximate analytical solution of the ABM, shedding light on the structure of the model and on the effect of the parameters. We then simulate the ABM and find that: (i) a more unequal income distribution lowers the prices globally, but implies stronger segregation; (ii) a spike of the demand in one part of the city increases the prices all over the city; (iii) subsidies are more efficient than taxes in fostering social mixing.}
 
\footnotetext[1]{Corresponding author: marco.pangallo@maths.ox.ac.uk. We thank seminar participants at WEHIA 2016, EAEPE 2016, ISCPIF, SCCS 2015, IRES Piemonte and INET Oxford, in particular Henri Berestycki, Allan Davids, J. Doyne Farmer, Michele Loberto, Mauro Napoletano, David Pugh and Matteo Richiardi, for helpful comments. We also thank Florian Artinger and Gerd Gigerenzer for pointing us to their work on aspiration level heuristics, and Marco Ranaldi, R. Maria del Rio Chanona and two anonymous reviewers for carefully reading the manuscript. A previous version of this manuscript circulated under the title ``Price formation on a housing market and spatial income segregation''.  This paper reuses material from the master thesis of MP at the University of Torino. MP especially thanks Pietro Terna, who was the main advisor of his thesis, for insightful and frequent discussions. }

\renewcommand*{\thefootnote}{\arabic{footnote}}

\vspace{20pt}

\textbf{J.E.L. codes}: C63, D31, R23, R31 

\textbf{Keywords:} Agent-Based Model, Housing Market, Spatial Segregation, Aspiration Level Heuristic, Income Inequality
\vspace{10pt}

\newpage

\section{Introduction}
\label{sec:intro} 

The allocation of people into the most productive cities is becoming an issue of central importance in a globalized world. For example, \cite{hsieh2017housing} calculate that U.S. GDP could be 36\% higher had workers been able to settle in the most productive areas.  Within cities, the place where people live and the way they are distributed has tremendous consequences on educational and labor opportunities \citep{benabou1993workings} and on the provision of public goods \citep{tiebout1956pure}. Spatial segregation of households along different income groups influences the wealth distribution and contributes to income inequality \citep{reardon2011income,rognlie2016deciphering}. A better understanding of the relation between segregation, income inequality and house prices and of the policies that are designed to deal with these issues is the main focus of this paper. 

From a theoretical perspective, the residential income distribution has traditionally been studied using spatial equilibrium models \citep{fujita1989urban,duranton2015urban}. These models assume that for each income group there exists a constant utility level across the city, and solve for the equilibrium prices yielding the same utility in each neighborhood. This framework allows for several non-trivial insights about the residential choices of fully optimizing agents. However, it does not explicitly represent the behavior of buyers and sellers and does not illustrate the market dynamics that lead some households to segregate in certain areas of the city (or to be segregated out of the city).

In this paper we build on a substantial body of work on Agent-Based Models (ABMs) of the housing market to introduce a parsimonious and tractable ABM specifically suited to study income segregation. Differently from spatial equilibrium models, the outcomes of our ABM result from unconstrained interactions between economic agents. Indeed, buyers and sellers follow simple behavioral rules and heuristics and their interactions determine prices and segregation patterns without the imposition of any aggregate constraint such as equilibrium. This is useful for various reasons.

On the one hand, explicitly representing the functioning of the housing market makes it possible to provide more realistic narratives than in spatial equilibrium models for a variety of phenomena. Consider for example the mechanism that segregates poor households out of the most attractive locations in a city, defined as the places with highest density of \textit{amenities}. In the spatial equilibrium model of \cite{brueckner1999central}, segregation occurs when the marginal valuation of amenities rises sharply with income, so that high-income households are willing to bid more than low-income households to reside in the place with most amenities.\footnote{\cite{brueckner1999central} assume that this place is the center of the city.} This means that the rich value amenities more than the poor do. In our ABM rich and poor households value amenities in the same way, but segregation is simply explained by the bidding process and by the market dynamics -- poor households would like to live in the most attractive places, but as the prices go up they cannot afford it.

On the other hand, modeling the housing market as the outcome of decentralized and unconstrained interactions makes it possible to fully consider various forms of heterogeneity. In this paper we focus on income heterogeneity among the buyers to study how income inequality shapes house prices and residential income segregation.  We show that higher income inequality worsens income segregation, but reduces the average level of the prices. We also study how subsidies and taxes may foster social mixing, showing that these policies have different effects on distinct parts of the income distribution. This reinforces the importance of heterogeneity in evaluating the effectiveness of different policies.

The building blocks of our ABM are the behavioral rules for buyers and sellers and the price formation mechanism. The buyers, who differ by income, have to select among heterogeneous neighborhoods where they search for a dwelling. Instead of maximizing their expected utility, the buyers select a neighborhood with a probability proportional to their utility, as in discrete choice theory \citep{anderson1992discrete}.\footnote{\cite{epple1998equilibrium} and \cite{Bayer2004} introduce discrete choice theory into spatial equilibrium models.} The sellers determine their reservation price -- the minimum price they are willing to accept -- by employing an \textit{aspiration level heuristic}. This means that sellers try to apply a markup on the market price at the time when they put their dwelling on sale, and progressively reduce their reservation price if their sale is unsuccessful.\footnote{This concept was first proposed by \cite{simon1955behavioral}, and successively developed in search theory \citep{wheaton1990vacancy,genesove2012search,han2015microstructure}.} Many models of the housing market assume that sellers act optimally conditional on the distribution of offers by potential buyers \citep{anenberg2016information}. Here we follow instead the \textit{fast and frugal heuristics} paradigm \citep{gigerenzer1999simple} and assume that sellers employ a fixed selling rule, without an explicit attempt to optimize profits \citep{artinger2016heuristic}. Given that in real housing markets information is limited and dispersed, using simple heuristics may in fact be optimal \citep{gigerenzer1999simple}. Finally, in our ABM buyers and sellers are matched through a continuous double auction taking place in each neighborhood at every time step. This represents bilateral bargaining in a stylized way. The market price in each neighborhood is simply the average of the prices of the transactions that occurred there.

We begin analyzing the model by finding an approximate analytical solution of the ABM. Our strategy is to start from the simplest setting -- e.g. all buyers have the same income -- and to increase the complexity of the model in a modular way. We do not attempt at finding a general solution, which would be unfeasible, but we provide insights in specific settings that are valid in the most complex settings too. From a methodological point of view, this approach provides two main contributions. First, we show that substantial simplifications can still capture some important aspects of an ABM while allowing for mathematical tractability, as in \cite{gualdi2015endogenous}. Second, the closed-form solutions provide insights that can be used for calibration. For instance, some parameters only occur as a combination (e.g. as a ratio of one parameter to the other), so it is sufficient to analyze the effect of one parameter while holding the other fixed. In applied mathematics and physics, these combinations are known as \textit{effective parameters}.

We then simulate the ABM and focus on the relation between residential segregation, income distribution of the buyers, house prices and policies. First, we find that stronger income inequality leads to stronger residential segregation, in accordance with empirical evidence \citep{reardon2011income}. To the best of our knowledge, our model is the first to account for this empirical fact. Interestingly, stronger income inequality also leads to lower average prices at the city level, as empirically confirmed too. \cite{maattanen2014income} explain this stylized fact by introducing a matching model with an optimal assignment rule between households acting as buyers and households acting as sellers. In our ABM, this finding is simply explained by the price formation mechanism. As income inequality increases, fewer richer households bid higher, while the majority of households bid lower. Because each buyer only bids for one dwelling and the market price at each location is the average of all transaction prices, the global effect is negative. 

Second, we find that a spike of demand in one part of the city increases the prices all over the city. We model the rise of demand as an additional influx of high-income households in the most attractive locations, mimicking the process of rich ``foreigners'' trying to purchase properties in the city \citep{chinco2015misinformed}. The interpretation for this finding is as follows. An increase of the prices in high-income neighborhoods implies that some of the households that would have considered moving there move instead to low-income locations. This is due to a substitution effect with the non-housing consumption good that buyers also consider in their utility function. Therefore, the prices increase at the least attractive locations too and the lowest-income households get segregated out of the city. \cite{favilukis2017out} come to the same conclusion within an overlapping generations model, and this finding gets some empirical support from \cite{cvijanovic2018real} and \cite{sa2016effect}. 

Third, we implement a system of \textit{ad-valorem} taxes and subsidies on buyers, and investigate which policy is most effective at fostering social mixing. Low-income households receive subsidies, whereas high-income households have to pay a buyer transaction tax. Effectively, these policies reduce the income spread between the households, but the effect is different if they target the low-end or the high-end of the income distribution. Subsidies directly target the low-income households and make it possible for them to buy properties in previously not affordable neighborhoods. Taxes on the contrary reduce the reservation prices of high-income households, but have no significant effect on transaction prices because transaction prices are still below the reservation prices of the rich. Note that in our model the only transmission channel for taxes is through reducing the reservation prices. As we assume inelastic global demand, taxes do not crowd out high-income households \citep{dachis2011effects}.\footnote{However, at least for the very top of the income distribution, demand is likely to be quite inelastic if houses are purchased as a primary residence. Taxes could instead discourage investors and secondary residence buyers.}  We finally perform a welfare analysis of the different policies, concluding that both subsidies and taxes increase the welfare of poor households at the expense of the welfare of rich households. The change in social welfare is negative if one considers a utilitarian welfare function, but it is positive according to a Rawlsian welfare function (see e.g. \citealt{mas1995microeconomic} for definitions of these functions).

We conclude this introduction relating our model to other ABMs of the housing market, and laying out a roadmap.

\paragraph{Relation to the literature on Agent-Based Models of the housing market.}

With respect to the rest of the literature, we designed our ABM to be parsimonious and tractable, yet with a sufficiently detailed description of the economic forces that shape the housing market. This choice was in part to ensure some mathematical tractability. However, we also wanted our agents to behave according to simple heuristics, as this may be optimal in situations in which information is limited and dispersed \citep{gigerenzer1999simple}, such as in housing markets. 

We start comparing our model to some of the earliest ABMs modeling housing markets and income segregation. \cite{feitosa2008spatial} show how segregation can emerge even if one considers the simplest setting with the minimal number of parameters. \cite{jackson2008agent} also introduce a relatively simple model, focusing on how the interaction between different classes of agents (``professionals'',``college students'', ``non-professionals'', and ``elderly'') may lead to gentrification. The opposite end of the complexity spectrum is taken by \cite{gilbert2009agent}: They build an ABM where some houses can be constructed, others are demolished, some agents may put their apartment on sale because they lost their job, etc. The complexity of our ABM is in between the contributions of \cite{feitosa2008spatial} and \cite{jackson2008agent} and that of  \cite{gilbert2009agent}. \cite{filatova2009agent} introduce an ABM with a similar complexity to the model in the present paper, but they assume that the reservation price of the sellers is just 25\% more of the so-called agricultural rent \citep{fujita1989urban}. Our aim here is to represent the supply side in a more detailed way. We mostly build on \cite{gauvin2013modeling}, by which our model shares many assumptions but differs substantially on the behavioral rules and the market mechanism. 

\cite{ettema2011multi} represents the supply side in a very detailed way. He assumes that sellers calculate the probability to sell at each list price, and choose the optimal expected sale price taking into account the disutility of a delay in selling. Moreover, sellers use Bayesian learning to update their perceived probabilities of selling. As mentioned above, in our ABM we favor a simpler heuristic to determine the list price. \cite{magliocca2011economic} represent the formation of a city through the interaction of land and housing markets. They highlight the importance of path dependence in determining the shape of the city. In our ABM we start with an existing city and focus on the long run prices and segregation patterns if the current state of the city was to persist. In this situation we do not find that path dependency plays a significant role. \cite{delloye2015morphology} also consider path dependency, but the shape of the city is determined by the strength of agglomeration economies. The ABMs in \cite{delloye2015morphology} and \cite{lemoy2017exploring} are specifically designed to potentially reach the equilibrium of spatial equilibrium models \citep{fujita1989urban}. In our ABM we do not have this goal, and even when the system reaches a steady state (punctuated by noise) this is not an equilibrium in the typical sense of urban economics.

\cite{huang2013effects} systematically study the effect of heterogeneity on the outcomes of the ABM. They focus on two types of heterogeneity: income heterogeneity and preference heterogeneity. In our ABM we also consider income heterogeneity. Preference heterogeneity follows from different budget constraints among buyers, but is not modeled explicitly by letting different agents have different parameters in their utility function. \cite{harting2018residential} study the relation between income heterogeneity and ethnic preferences, finding complex non-linear relations among these variables. In particular, they find that subsidies may worsen segregation if they alter the delicate balance between income and ethnic preferences. This effect is absent in our model as we do not have ethnic preferences, and we find that subsidies have the most positive effect on social mixing. It should also be noted that the ABM of \cite{harting2018residential} is not spatially explicit, while our ABM is.

A final dimension in which our model relates to the rest of the literature is the use of data to micro-calibrate the agents and the parameters. \cite{geanakoplos2012getting} and 
\cite{baptista2016macroprudential} use a wealth of microdata to calibrate non-spatial housing market ABMs and study implications on systemic risk, while \cite{filatova2015empirical} and \cite{de2017bridging} introduce geographically detailed features in their spatial ABMs through the inclusion of GIS information. Our goal here is not to calibrate our model to a specific city or country, although we do calibrate one parameter on real data as it is stable across different datasets (see Appendix \ref{sec:parvalues}). 

\paragraph{Roadmap.} The rest of this paper is organized as follows. In Section \ref{sec:model} we present the model; we then give an approximate mathematical description of it in Section \ref{sec:math}, and provide results from the numerical simulations in Section \ref{sec:num}. Section \ref{sec:conclusion} concludes.

\section{Model}
\label{sec:model} 

A schematic representation of the model is provided in Figure \ref{fig:figmodel}, and an ODD+D \citep{grimm2010odd,muller2013describing} description of the ABM is given in Appendix \ref{sec:oddd}. Our model considers an already formed city, mathematically defined as a grid in the Cartesian plane. A specific point of the grid is a \textit{location}. Space is characterized by different levels of \textit{attractiveness}, a variable subsuming exogenous intrinsic features and endogenous social characteristics. Time is discrete. At each time step:
\begin{enumerate}
\item Some households -- the buyers -- come to the city from outside and try to purchase a dwelling in the metropolitan housing market. They select a location with a probability proportional to their expected utility at that location. 
\item Households already living in the city decide to put their dwelling on sale with a certain probability. We refer to the households with a dwelling on the market as ``the sellers''.
\item The buyers have heterogeneous incomes and \textit{bid} a certain amount in order to secure a property. The bids are proportional to their income. 
\item The sellers determine the price they \textit{ask} by employing an aspiration level heuristic. 
\item At each location, buyers and sellers are matched through a continuous double auction. The transaction price is a weighted average of the bid and ask prices. The weight depends on the bargaining power.
\item Successful buyers take residence in the location where they searched and successful sellers leave the city. The market price is computed at each location as the average price of the transactions that occurred there.
\end{enumerate}

The goal of this paper is to introduce a \textit{baseline} ABM of the housing market, in which realistic behavioral rules give rise to a variety of phenomena related to income inequality and income segregation. Therefore we simplify many aspects of the model, which is also necessary to keep the ABM partly analytically tractable. While describing the model, we highlight a number of possible modifications that would improve the realism of the ABM. Most importantly, in this paper we only focus on steady states -- our model describes what would happen to the spatial price distribution and income segregation patterns in the long run, if the current state of the city was to persist. 

\begin{figure}[!ht]
\centering
\includegraphics[width=.7\textwidth]{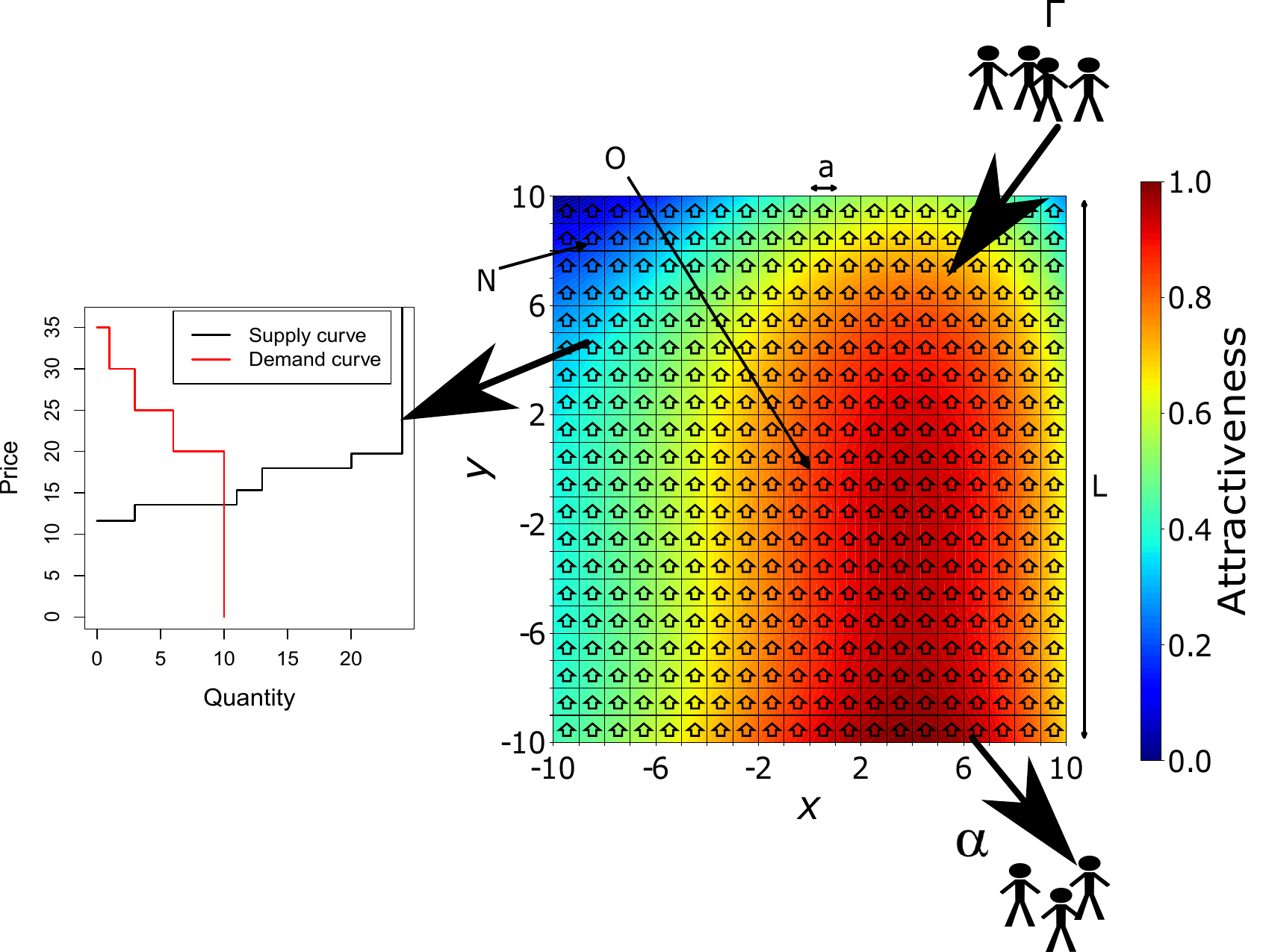}
\caption{\textbf{Schematic representation of the model.} The city is a grid of \textit{locations} $X=(x,y)$, with center $O$, linear size $L$ and distance $a$ between locations. At each location there are $N$ identical housing units, and space is characterized by different levels of \textit{attractiveness}. (The attractiveness evolves over time according to the social composition of the neighborhood, in this schematic we show the attractiveness at a particular time step.) At every time step, $\Gamma$ buyers come to the city from the outside and current inhabitants put their dwelling on sale with probability $\alpha$. At each location and every time step, buyers and sellers are matched through a continuous double auction, here visually depicted with discrete demand and supply curves constructed from the reservation prices of the agents.}
\label{fig:figmodel}
\end{figure}

\subsection{Space and time}
\label{sec:model1} 

The city is defined as a two-dimensional square grid $\Omega$ of locations $X \in \mathds{Z}^2$, with linear size $L$. The origin $O$ is taken as the geographical center of the city, and two neighboring locations are separated by a distance $a$. The same number $N$ of identical dwellings\footnote{Housing market segmentation can be important \citep{landvoigt2015housing,piazzesi2015segmented}. Moreover, apartments are at least heterogeneous for what concerns their size. We experimented with heterogeneous size and only found a blurring of the segregation patterns. For instance, assuming that with homogeneous size the rich are completely segregated in the center, with heterogenous size a few low-income households afford to reside in the center in small apartments.} are available at each location $X \in \Omega$. Time is discrete and indexed by $t$. The time horizon is infinite.

All locations are characterized by an attractiveness $A(X,t)$. This combines a constant and exogenous intrinsic attractiveness $A^0(X)$ and an endogenous social component $A^S(X,t)$. The intrinsic attractiveness quantifies the presence of natural amenities, historic buildings, and the convenience of the transportation system at that location.\footnote{The typical way to include transportation into urban economics models is through a cost that is included in the budget constraint \citep{fujita1989urban}. However, especially in European cities, transportation costs are relatively small, and so we model the commuting time as a disutility -- in particular as a negative effect on the attractiveness. \cite{gaigne2018amenities} show that (within the traditional modeling framework) amenities and commuting costs can be aggregated into a single quantity, which they name the \textit{location-quality index}.} The social component quantifies the educational and labor opportunities that arise from living in that neighborhood \citep{benabou1993workings}. These in turn depend on the average income of the agents living in that neighborhood, so we assume that $A^S(X,t)=\overline{Y}(X,t)/ \overline{Y}(t)$, where $\overline{Y}(X,t)$ is the average income of the agents living at location $X$ at time $t$, and $\overline{Y}(t)$ is the average income of the agents in the city at the same time step. We finally assume that the intrinsic and social components combine multiplicatively, namely 

\begin{equation}
A(X,t)=A^0(X) A^S(X,t).
\label{eq:totalattractiveness}
\end{equation}

One remark is important here. For simplicity in this paper we only consider a monocentric city where the intrinsic attractiveness $A^0(X)$ decreases with the distance from the center (differently from Figure \ref{fig:figmodel}). As we will see, this creates a positive feedback by which rich agents take residence in the center and further increase the total attractiveness $A(X,t)$ there through an increase of the social component $A^S(X,t)$.\footnote{Within our model the social component has a minor effect with respect to the intrinsic component. See Section \ref{sec:dynamics}.} This captures the idea of a city where most amenities and job opportunities are concentrated in the center \citep{brueckner1999central}, also normally called Central Business District (CBD) in urban economics \citep{fujita1989urban}. By tuning $A^0(X)$, our ABM trivially generalizes to cities with arbitrarily distributed attractiveness -- as in Figure \ref{fig:figmodel} -- including polycentric cities \citep{fujita1982multiple,lucas2002internal}, cities where the rich choose to live in the peripheries \citep{alonso1964location}, or cities with any other shape as studied in urban geography and city planning (see e.g. \citealt{angel2016spatial}).

\subsection{Agents}
\label{sec:agents} 

The agents are households. At each time step $t$ a constant number $\Gamma$ of households arrive on the market from outside the city and try to purchase one property. We refer to these agents as ``the buyers''. These households are only characterized by their monthly income $Y$. For simplicity, we consider a finite number $K$ of income levels. Agents with the same income are denoted by $k$-agents, $k\in \{1,...,K\}$, and have income $Y_k$. These incomes are ordered by increasing values, $Y_1 < Y_2 <... <Y_{K}$, and are separated by a constant $\Delta$. We denote the number of incoming agents in each income category by $\Gamma_k$, s.t. $\sum_{k}{\Gamma_k} = \Gamma$. 

At the end of each time step, some of the buyers secure a property and take residence in the city -- we refer to these agents as ``housed''. We assume that the unsuccessful buyers leave the city and may come back with a subsequent cohort. Indeed, it is not useful to keep track of the identities of the buyers, as these agents are only characterized by their income.

At the beginning of each time step, households already living in the city may put their dwelling on sale with a constant and homogeneous probability $\alpha$. Housed agents whose dwelling is on the market are denoted as ``sellers''. The sum of buyers, sellers and housed agents is constant and equal to $\Gamma+NL^2$. At $t=0$, all agents in the city are housed. When $t>0$, the relative proportion of housed and sellers depends on the number of buyers $\Gamma$, on the sale probability $\alpha$ and on the market outcomes.

In this model $\Gamma$ and $\alpha$ are fixed, but it could be useful to make them depend on the level of the prices in the city and on the timing with respect to the house price cycle. For instance, $\alpha$ could be larger if the prices are rising and smaller if the prices are falling.  However, being primarily concerned with income segregation, we are mostly interested in the long-term trends and the relative prices between several areas of the city, which are influenced more strongly by the attractiveness than by the global demand or leave rate. 

\subsection{Demand}
\label{sec:demand}

At each time step $t$, demand at each location $X$ is determined by the individual decisions of the buyers. The $k$-buyers have utility function $U_k(X,t) = z_k(t)^{1-\beta} \left(A(X,t) \right)^\beta$, where $z_k(t)$ represents the monthly non-housing consumption of $k$-agents at time $t$. (A very similar utility function is very common in urban economics, see \citealt{fujita1989urban}.) The parameter $\beta \in [0,1]$ is the weight given to the attractiveness, which exemplifies housing consumption. The budget constraint is
\begin{equation}
z_k(t) + P(X,t) = Y_k,
\label{eq:bc}
\end{equation}
where $P(X,t)$ is the expected monthly mortgage repayment after purchasing a property at location $X$ and time $t$ (the determination of $P(X,t)$ will be explained at the end of Section \ref{sec:model}). We have also assumed unit cost for the non-housing consumption good, which is the num\'eraire of the economy. (This is a standard modeling choice in urban economics.) 
The above equation does not assume savings. If we wanted to include savings, we could just consider that the buyers spend a fraction $s$ of their income $Y_k$, and redefine the incomes in the model as the part that is not saved.  Replacing the budget constraint in the utility function, we get the indirect utility
\begin{equation}
    V_k(X,t)=
\begin{cases}
    \left( Y_k - P(X,t)  \right)^{1-\beta} \left(A(X,t) \right)^\beta,& Y_k > P(X,t),\\
    0,              & Y_k \leq P(X,t).
\end{cases}
\label{eq:utilityf}
\end{equation}
In case $Y_k \leq P(X,t)$ the $k$-agents cannot afford to purchase a property at location $X$ and time $t$, and so their utility would be $0$. Note that this is not an indirect utility function standard in microeconomics (see e.g. \citealt{mas1995microeconomic}). The reason is that $A(X,t)$ is not a good whose quantity to be purchased is determined optimally, but rather a fixed quantity that is taken as given.

We consider a probabilistic model for the choice of residential location by the buyers. The $k$-buyers choose location $X$ at time $t$ with probability $\pi_k(X,t)$, proportional to the utility they expect to find at that location: 
\begin{equation}
\pi_k(X,t)=V_k(X,t)/{\sum_{X'\in \Omega}  V_k(X',t)}. 
\label{eq:choiceprob}
\end{equation}
This behavioral rule follows the literature on discrete choice theory \citep{anderson1992discrete}, and captures the idea that the decisions of the buyers are noisy, although not irrational.\footnote{Typically, in discrete choice theory the probability $\pi_k(X,t)$ depends exponentially on the utility $V_k(X,t)$, with a sensitivity parameter -- the \textit{intensity of choice} -- that quantifies the deviation from rationality. Here we choose a polynomial form of degree one, which is a more parsimonious parametrization.} 

To sum up, the demand side of the market at location $X$ and time $t$ is characterized by the number of $k$-buyers $N_b^k(X,t), \forall k$ -- determined stochastically from Eq. \eqref{eq:choiceprob} -- and by their reservation demand price, which we simply assume to be a multiple of their monthly income: $P^d_k = \zeta Y_k$. The reservation demand price is the maximum amount the buyers are willing to bid. So they may potentially borrow an amount $\zeta Y_k$ from a bank, and repay it monthly in $\zeta$ installments. For clarity of exposition, we will consider $\zeta = 1$, but this is equivalent to normalizing all prices to their monthly equivalent.
 
\subsection{Supply}
\label{sec:supply}

The dwellings available for sale at location $X$ and time $t$ are those that are put on the market by the agents housed in $X$ in the same time step plus, if any, those that have not yet been sold on previous time steps. We denote the number of sellers as $N_s(X,t)$.

The reservation offer price is the minimum amount a seller is willing to accept to sell his property. The sellers determine their reservation price by employing an \textit{aspiration level heuristic}. This concept was initially proposed by \cite{simon1955behavioral}, who in fact specifically considered the example of an individual trying to sell a dwelling. The seller would accept any offer above a \textit{satisficing} threshold, and adjust that threshold downward if the sale was unsuccessful. Search theory has developed this idea by endogenizing the threshold and the waiting time so to follow an optimal stopping rule. A rich literature has applied this idea to model the time on market of real-estate properties. 

Most models in search theory assume that the sellers know the distribution of offers by the potential buyers. In real housing markets information is limited and dispersed, and in our model there is an inherent stochasticity due to the noisy decisions of the buyers. \cite{anenberg2016information} introduces a model in which sellers update their reservation price using Bayesian learning on the received offers, so to behave optimally given the available information. However, in uncertain environments the use of simple heuristics in place of optimization can be optimal. Therefore, following the \textit{fast and frugal heuristics} paradigm proposed by \cite{gigerenzer1999simple}, we assume that the sellers employ a fixed selling rule, without an explicit attempt to optimize profits. In particular, they try to apply a markup on the market price, and progressively reduce their reservation price as their sale is unsuccessful. \cite{artinger2016heuristic} show empirically that most sellers indeed follow this behavioral rule, which yields more profits than equilibrium strategies.\footnote{The authors actually consider the used cars market and not the housing market, but the information structure is quite similar in the two settings.}

We follow \cite{artinger2016heuristic} in specifying the functional form of the aspiration level heuristic. Sellers first attempt to sell at a higher price than the market price, trying to apply a markup $\mu$ on the current market price. If the sellers are unsuccessful in selling after $\tau$ time steps, they adjust their reservation price downward by a factor $\lambda$. Here $\lambda$ captures the ``downward stickiness'' typical of housing market: when demand increases, prices rise quickly, but in situations of excess supply prices decrease slowly. In formula, the reservation offer price $P^s_i(X,t)$ for seller $i$ at location $X$ at time $t$ is
\begin{equation}
P^s_i(X,t) = (1+\mu)P(X,t_i)\lambda^m,\text{ with } m\tau \leq t - t_i < (m+1)\tau,
\label{eq:alh}
\end{equation}
where $t_i$ is the time step in which household $i$ put their apartment on sale, and $t-t_i$ is the time on market. The reservation price is decreased by a factor $\lambda$ every multiple $m$ of $\tau$. Note that -- differently from the case of the buyers -- here we need to keep track of the identities of all sellers $i$, as all reservation offer prices can potentially be heterogeneous.\footnote{Here we do not assume heterogeneity in the parameters $\mu$, $\lambda$ and $\tau$. We do so for various reasons. On a practical note, we experimented with having these parameters drawn from a distribution and virtually nothing changed at the aggregate level. As we  aim to keep our model as parsimonious and tractable as possible, it makes sense to just assume homogeneity of these parameters. On a theoretical note, \cite{artinger2016heuristic} show in their Figure III (p.~25) that the distribution of these parameters (that they call $\alpha$, $\gamma$ and $\beta$ respectively) is rather peaked around the median values in the used cars market they analyze, and we think that this may also apply in the housing market.} 

\subsection{Matching}
\label{sec:matching}

Matching between buyers and sellers occurs at each location and each time step through a continuous double auction, which we take as a stylized model of a bilateral bargaining process. The reservation offer (demand) prices enter in a random sequence as asks (bids) in a limit order book. Every time a bid price exceeds an ask price, a transaction takes place and the two prices are removed from the order book. This process continues until all agents have placed their orders. 

The price of the transaction depends on bargaining, which we model in a stylized fashion. In particular, we assume that the price of a transaction between a $k$-buyer and seller $i$ is a linear combination of the reservation prices, 
\begin{equation}
P_{ki}=\nu P_k^d + (1-\nu)P_i^s.
\label{eq:marketprice}
\end{equation}
Here $\nu$ quantifies the bargaining power of the seller. If $\nu=0$, the transaction price is simply the reservation price of the seller -- this parameterization would model a situation in which the seller needs to post his reservation price -- while if $\nu>0$ the seller would post a higher price. 

Finally, the market price $P(X,t)$ is the average of all transaction prices that were recorded at location $X$ and time $t$.

\section{Mathematical analysis}
\label{sec:math}

The goal of this section is to show that, in spite of the complexity of the ABM, some of its features can be understood analytically, without the need to resort to numerical simulations. Our analytical solution gives insights on the causal mechanisms of the ABM and makes the effect of its parameters clearer. While our mathematical analysis gives some insights into the determinants of segregation, the reader who is mostly interested about the relation between segregation and income inequality and in the effect of subsidies and taxes may skip this section and move to the numerical simulations that follow in Section \ref{sec:num}.

For our analysis we follow a modular strategy, in the sense that we progressively focus on specific aspects of the ABM while neglecting other features in order to maintain tractability. In particular, in Section \ref{sec:M1} we assume that the agents only value the attractiveness in their utility function (i.e. $\beta=1$ in Eq. \ref{eq:utilityf}) and we consider only one income category (i.e. $K=1$). In Section \ref{sec:M2} we relax the assumption that $\beta=1$, and focus on the tradeoff between non-housing consumption and attractiveness. In Section \ref{sec:M4} we consider two income categories (i.e. $K=2$) and study the conditions that imply income segregation. 

In this section we only report the details necessary to understand the mathematical analysis, while leaving some technical derivations to Appendix \ref{sec:mathderivations}.

\subsection{Preliminary steps}
\label{sec:math1} 

The following simplifying assumptions are made for analytical tractability.  These simplifications are then relaxed in the numerical simulations. 

First, we average out stochastic effects by taking expected values. For example, although the number of buyers coming to each location $X$ is given in the ABM by a multinomial stochastic process with probabilities defined in Eq. \eqref{eq:choiceprob}, for the mathematical analysis we assume that the number of buyers at $X$ is the expected value of the process. 

Second, we assume continuous space by considering a vanishing distance between locations, $a \rightarrow 0$. This step requires an important technical attention. So far all quantities were defined for each location $X$ -- e.g. the number of $k$-buyers at $X$ was $N_b^k(X,t)$. In order to take the continuum limit we cannot keep using this definition. Indeed, when $a \rightarrow 0$ and $L$ is fixed the number of locations ($L/a$) grows large, so for example the number of $k$-buyers in any specific $X$ would become vanishingly small (for fixed total number of buyers $\Gamma$). We solve this technical issue by dividing all quantities that are defined at the location level by the local area $a^2$. These quantities are now mathematically defined as \textit{densities}, and we denote them by a lower-case letter. For example the density of $k$-buyers at $X$ and $t$ is $n_b^k(X,t)=N_b^k(X,t)/a^2$. When $a \rightarrow 0$, also $N_b^k(X,t) \rightarrow 0$, and so the density is well defined in the limit. The other variables we have to transform are $N_s(X,t) \rightarrow n_s(X,t) = N_s(X,t)/a^2$ and $N \rightarrow n = N/a^2$, where $n=n(X)$ is uniform for all $X$.

Third, we neglect the social component of the attractiveness, making the attractiveness time independent: $A(X,t) \rightarrow A(X)$. This is exact in the case of one income category, because all agents have the same income and so $\overline{Y}(X,t)=\overline{Y}(t)$ and $A(X,t)=A^0(X)$ (see Eq. \ref{eq:totalattractiveness}).\footnote{We will also show that ignoring the social component is safe in the case of two income categories too, at least within our mathematical analysis (see Section \ref{sec:M4}).}  We further assume that the intrinsic component of the attractiveness has circular symmetry and decreases with the distance $r$ from the center $O$ (monocentric city), up to a radius $R_{max}$ that represents the borders of the city. Therefore, we can write the attractiveness as $A(r)$, with $A'(r)<0$. A possible specification of $A(r)$ is
\begin{equation}
    A(r)=
\begin{cases}
    e^{-\frac{r^2}{R^2}},& 0 < r \leq R_{max},\\
    0,              & r> R_{max},
\end{cases}
\label{eq:intrattr}
\end{equation}
where $R$ is a steepness parameter that quantifies how much the attractiveness is concentrated in the center.\footnote{Here $R$ plays the same role as the intensity of choice in discrete choice models. This is another reason why we choose the polynomial form in Eq. \eqref{eq:choiceprob} instead of the more common logit form.} If $R$ is very small, only the center is very attractive and the peripheries are not attractive; if $R$ is large, the attractiveness is spread evenly across the city. $R_{max}$ is chosen so that the areas of the discrete-space square lattice $\Omega$ and its continuous-space circular approximation considered in Eq. \eqref{eq:intrattr} are the same, that is $\pi R_{max}^2 = L^2$. In order to emphasize that the results that follow do not depend on the specific form of the attractiveness in Eq. \eqref{eq:intrattr}, we write in general $A(r)$, except when performing specific calculations.

Finally, we focus on the steady state of the model. As already mentioned, our ABM determines the economic outcomes in the long run, if the current state of the city was to persist. It then makes sense to consider time averages in the numerical simulations, and we will show that these closely match the analytical steady states.

\subsection{Baseline case}
\label{sec:M1}

Here we only consider one income category. We also assume that the buyers only value the attractiveness in their utility function (and so do not value the non-housing consumption). We only sketch the derivation and discuss which assumptions are made, while the full derivation is given in Appendix \ref{sec:mathderivationsonecat}.

Our first step is to compute the densities of buyers and sellers at any distance $r$ in the steady state. Calculating the density of buyers is trivial, as it only depends on the attractiveness which is exogenous in the simplification considered in this section. Calculating the density of sellers is instead tricky, because it depends on how many dwellings have remained on sale from the previous time step. It is possible to find a self-consistent expression for this density for locations at distance $r$ if one assumes that all buyers were successful in securing an apartment at the same locations in the previous time step. As we will see, breaking of this condition leads to inconsistencies and gives insights into the functioning of the model. The second step of our derivation involves computing the reservation prices of buyers and sellers. All buyers have the same reservation price -- we are assuming only one income category. We calculate the expected reservation price of the sellers given the expected time it takes to sell a dwelling at a location at distance $r$. We finally compute the market price as a weighted average of the reservation prices of buyers and sellers.  

The steady state market price $P^\star(r)$ at distance $r$ reads 
\begin{equation}
P^\star(r)=\frac{\nu Y \left( n - \frac{1-\alpha}{\alpha}\Gamma\frac{A(r)}{Z}-\lambda^{1/\tau}\left( n -\frac{\Gamma}{\alpha}\frac{A(r)}{Z} \right)\right)}{\left( n - \frac{1-\alpha}{\alpha}\Gamma\frac{A(r)}{Z}-\lambda^{1/\tau}\left( n -\frac{\Gamma}{\alpha}\frac{A(r)}{Z} \right)\right)-(1-\nu)(1+\mu)\Gamma\frac{A(r)}{Z}},
\label{eq:mathfixedpricealpha}
\end{equation}
where $Z=2\pi \int_0^{R_{max}}{r A(r) dr}$ is a normalization factor.
As mentioned above, this expression is correct provided that all buyers are always successful in securing an apartment, which in turn depends on two conditions being satisfied:
\begin{equation}
\frac{n_b^\star(r)}{n_s^\star(r)} \leq 1 \text{ and } P^\star(r) < Y.
\label{eq:conditions}
\end{equation}
The first condition means that there must be at most as many buyers as sellers; the second conditions states that the market price must be smaller than the income of the buyers. We now look at three limiting cases that are interesting in their own right and in which the conditions \eqref{eq:conditions} may or may not be satisfied. This also gives insights on the mechanisms of the model.

\textbf{Attractiveness.} We consider the least attractive locations, in which almost no buyers purchase any property. From Eq. \eqref{eq:mathfixedpricealpha}, $\lim_{A(r)/Z \rightarrow 0} P^\star (r) = \nu Y$. The reservation prices of the sellers drop to zero because the time on market grows to infinity. So from Eq. \eqref{eq:marketprice} the market price is simply determined by the bargaining parameter $\nu$ and by the reservation price of the buyers $Y$. The conditions \eqref{eq:conditions} are satisfied. 

\textbf{Tightness.} We consider some attractive location in which the number of buyers equals the number of sellers -- in other words the demand \textit{tightness} $q^\star=n_b^\star/n_s^\star=1$. We have $\lim_{q^\star \rightarrow 1} P^\star(r) = \nu Y / \left( 1 - (1+\mu)(1-\nu) \right)$. The expression for the market price simplifies considerably. However, it is possible to check that $\nu Y / \left( 1 - (1+\mu)(1-\nu) \right) \geq Y$, which is inconsistent with the assumptions of the model (sellers cannot bid more than their income $Y$). This is easily explained in dynamical terms. Suppose that at time $t$ the price is $P<Y$. The buyers bid $Y$, while the sellers ask $(1+\mu)P$. So the price at time $t+1$ is $P'=\nu Y + (1-\nu) (1+\mu)P > P$.  The sellers immediately sell their property, so their reservation offer price has no time to decrease, therefore a steady state cannot be reached as the market price keeps increasing. 

\textbf{Stickiness.} We now assume that the prices are extremely sticky, in the sense that the discount factor of the sellers $\lambda^{1/\tau}$ is close to unity. In the limit, 
$\lim_{\lambda^{1/\tau}\rightarrow 1} P^\star(r) = \nu Y / \left( 1 - (1+\mu)(1-\nu) \right)$. The outcome is the same as for $q^\star \rightarrow 1$. Indeed, the dynamics are similar. Assume that the agents put their dwelling on sale at $t$, when the price is $P$. At $t' \gg t$ they still ask $(1+\mu)P$, and so the market price at $r$ keeps increasing and cannot reach a steady state. Note that in Eq. \eqref{eq:mathfixedpricealpha}, $\lambda$ and $\tau$ occur as a combination of parameters, $\lambda^{1/\tau}$. They are effectively only one parameter, as it is enough to vary one while holding the other fixed.

Given the above analysis, we impose the constraint that if $P^\star(r)>Y$, we set $P^\star(r) = Y$. This constraint simply formalizes the idea that the buyers cannot bid more than $Y$, and so the market price cannot be higher.

\begin{figure}
\centering
\includegraphics[width=0.8\textwidth]{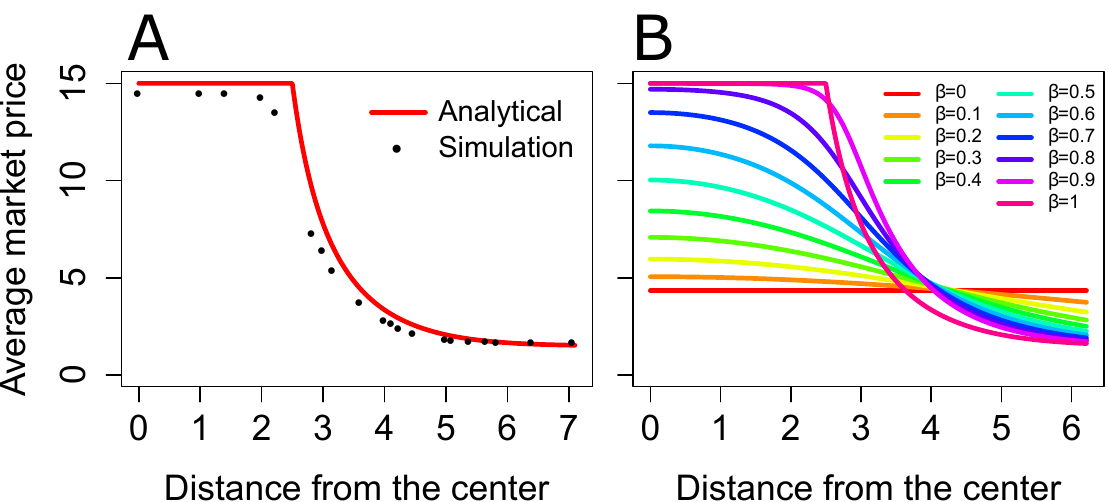}
\caption{\textbf{Analytical results with one income category.} (A) Comparison between the analytical solution (line) and the numerical simulations (dots), in the baseline case where $\beta=1$. The analytical prices are from Eq. \eqref{eq:mathfixedpricealpha}, setting $P^\star(r) = Y$ near the center where the conditions \eqref{eq:conditions} are not satisfied. The numerical results (here and in all simulations that follow) are averaged over 100 time steps over all locations at the same distance $r$, after an equilibration time of 50 time steps. We do not report standard errors on the numerical results because our model is ergodic and gives virtually identical results for different random seeds. (B) Effect of $\beta \in [0,1]$. The prices are more uniform if $\beta$ is small, i.e. the agents substitute housing with non-housing consumption. The other parameters are specified in Appendix \ref{sec:parvalues}.}
\label{fig:prices-analytical}
\end{figure}

As can be seen in Fig. \ref{fig:prices-analytical}A, the mathematical results are in line with the numerical simulations of the discrete-space and fully heterogeneous dynamics. The prices are slightly overestimated by the mathematical analysis. This is because of the order book dynamics. Sellers with a lower reservation price have higher chance of selling, but this effect is not captured in our mathematical analysis as we assume that all dwellings on the market at $r$ are sold with the same probability.

\subsection{General utility}
\label{sec:M2}

In general, the utility function for $k$-buyers for choosing a location at distance $r$ from the center at time $t$ is
\begin{equation}
U_k(r,t) = z_k(t)^{1-\beta} \left(A(r) \right)^\beta,
\label{eq:utility}
\end{equation}
where $0 \leq \beta \leq 1$. Since just the utility function is different from the previous section, the analysis is similar. (The only technical difference is described in Appendix \ref{sec:genutility}). As can be seen in Fig. \ref{fig:prices-analytical}B, for small values of $\beta$ the prices are almost uniform across the city, due to the substitution effect between housing and non-housing consumption, whereas larger values of $\beta$ increase the slope of the price gradient.

\subsection{Two categories}
\label{sec:M4}

\begin{figure}
\centering
\includegraphics[width=1.\textwidth]{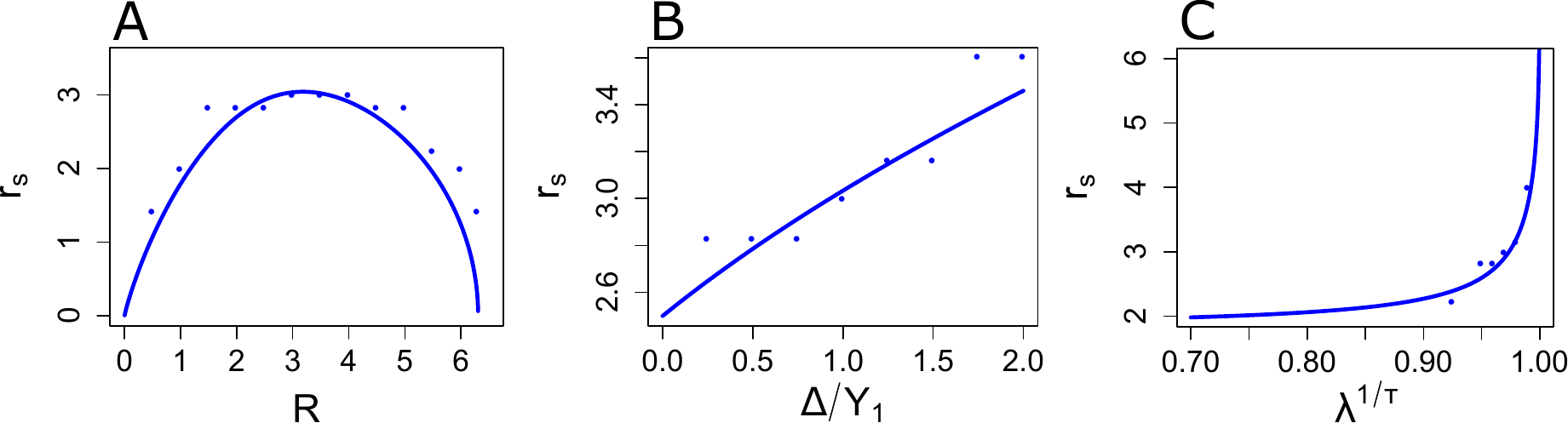}
\caption{\textbf{Size of the segregated region as a function of some parameters.} We plot the radius of the segregated region $r_s$ (Eq. \ref{eq:condsegr}) against three parameters. We compare the analytical results (lines) with numerical results (dots). In the numerical approximation we cannot have total segregation because of stochastic effects, so we define $r_s$ as the first distance at which the share of 2-agents is smaller than 95\%. (Note that in the numerical simulations $r_s$ takes only a finite number of values.) (A) $R$ quantifies how evenly spread is the attractiveness. (B) $\Delta/Y_1$ is the relative spread of the income levels. (C) $\lambda^{1/\tau}$ is the stickiness of the prices. The other parameters are specified in Appendix \ref{sec:parvalues}.}
\label{fig:segregation-analytical}
\end{figure}

Now we keep $\beta=1$ and consider two income categories, $K=2$. The income levels of the agents are $Y_1$ and $Y_2$, with $Y_2=Y_1+\Delta$. We denote the households respectively as \textit{1-agents} and \textit{2-agents} -- the poor and the rich respectively. We study under which conditions the market mechanism implies income segregation. Under the assumptions made in this section, the best way to determine this is to check if there exists a completely segregated circle with center $O$ and radius $r_s \in [0,R_{max}]$, which is only inhabited by 2-agents. In this circle the steady state value for the market price must be higher than the income level of the 1-agents, formally $P^\star(r)>Y_1, \; \forall r < r_s$. Since the circle is inhabited only by 2-agents, we can use the one category result in Eq. \eqref{eq:mathfixedpricealpha} with $Y=Y_2$ and $\Gamma=\Gamma_2$. Rearranging the boundary condition $P^\star(r_s)=Y_1$, and considering the specific form of the attractiveness as in Eq. \eqref{eq:intrattr}, we can compute $r_s$, the radius of the segregated region:
\begin{equation}
r_s=R\sqrt{\log{\frac{\Gamma_2 \left[ \frac{1}{1+\Delta/Y_1} \left( \frac{1}{\alpha} \left( 1-\alpha+\lambda^{1/\tau} \right)+ (1+\mu)(1-\nu) \right) - \frac{\nu}{\alpha} \left( 1-\alpha - \lambda^{1/\tau} \right) \right]}{\pi R^2 \left(1-e^{-R_{max}^2/R^2}\right) n \left( 1-\lambda^{1/\tau}\right) \frac{1-\nu-\nu\Delta/Y_1}{1+\Delta/Y_1}}}}.
\label{eq:condsegr}
\end{equation}
We can then study the effect of the parameters on $r_s$. The radius of the segregated region increases with the number of incoming high-income households $\Gamma_2$, and with the markup $\mu$ that the sellers try to apply. Indeed, $\Gamma_2$ is only in the numerator and multiplies the squared brackets, and $\mu$ is only in the numerator and provides a non-negative contribution. On the contrary, the radius $r_s$ decreases with the number of available apartments $n$ -- which is only in the denominator --, supporting the policy that less regulatory constraints on constructions may alleviate income segregation. The role of the other parameters cannot be immediately seen from Eq. \eqref{eq:condsegr}, so in Fig. \ref{fig:segregation-analytical} we plot the radius of the segregated region as a function of a few interesting parameters.

In Fig. \ref{fig:segregation-analytical}A we consider $R$, the scale factor of the exponential in Eq. \eqref{eq:intrattr} that quantifies how evenly spread is the attractiveness. Interestingly, as $R$ increases there is a non-monotonic effect. When $R$ is small (up to $R=3$ in Fig. \ref{fig:segregation-analytical}A), all 2-buyers want to reside in the center and this keeps the prices above the income level $Y_1$ of the 1-agents. The radius $r_s$ grows with $R$, because the 2-buyers spread in a larger region. But after a turning point, $r_s$ starts decreasing, because the 2-buyers spread more and more evenly and are not numerous enough at any specific location to keep the prices above $Y_1$. This result is in line with \cite{gaigne2018amenities}, who find that a multimodal distribution of amenities may foster social mixing. In Fig. \ref{fig:segregation-analytical}B we look at $\Delta/Y_1$, that quantifies the relative spread of income levels. These parameters always occur as a ratio of one another, suggesting that they can be treated as a unique parameter. A larger $\Delta/Y_1$ slightly increases segregation, almost in a linear fashion. Finally, in Fig. \ref{fig:segregation-analytical}C we show that more stickiness leads to more segregation. There is an asymptote at $\lambda^{1/\tau} \rightarrow 1$, in which the reservation offer prices never decrease, as discussed in Section \ref{sec:M1}.

In all the cases above analytical results match numerical results well, indicating that in this case it is safe ignoring the social component of the attractiveness (the segregated region is only inhabited by 2-agents).

\section{Numerical simulations}
\label{sec:num}

In this section we perform some numerical simulations of the fully-fledged ABM to analyze the effect of the income distribution and of subsidies and taxes on the prices and segregation patterns. We consider ten income categories, $K=10$, and we assume that the agents face a tradeoff between housing and non-housing consumption ($\beta=0.5$). The other values for the baseline parameters are discussed in Appendix \ref{sec:parvalues}. We set some parameters following guidance from the mathematical analysis, and set other parameters to empirically reasonable values. 

The cost of considering a more realistic setting is that no analytical solution is possible. To ensure full reproducibility, the code used to generate all figures is available at \texttt{https://dx.doi.org/10.5281/zenodo.1453347}.

\subsection{Dynamics of the Agent-Based Model}
\label{sec:dynamics}

\begin{figure}[!ht]
\centering
\includegraphics[width=1.\textwidth]{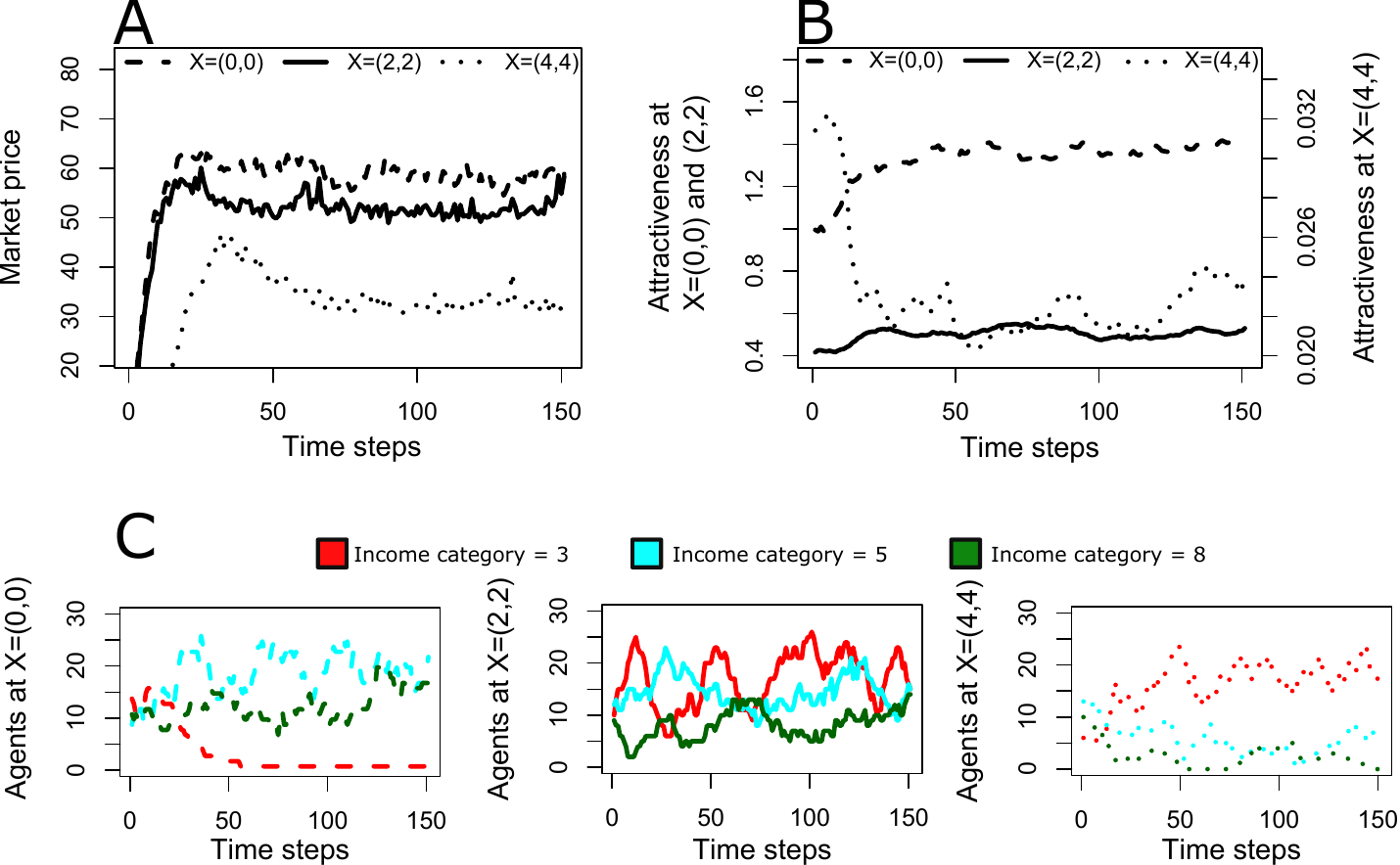}
\caption{\textbf{Dynamics of the Agent-Based Model.} We plot some variables over time at three locations representative of the city. $X=(0,0)$ is the center (dashed lines), $X=(4,4)$ is in the periphery (dotted lines), $X=(2,2)$ is in between (solid lines). We only show one simulation run, but the time series are qualitatively similar for other random seeds. Here $Y_1=30$ and $\Delta=11.86$, the other parameters are specified in Appendix \ref{sec:parvalues}. (A) The market price quickly reaches a steady state punctuated by noise. (B) The attractiveness $A(X,t)$ also reaches a similar state. At $X=(0,0)$ and $(2,2)$ (left scale on the vertical axis) it increases from the initial intrinsic value $A^0(X)$, whereas at $X=(4,4)$ (right scale) it decreases. (C) Number of agents with $k=3$ (poorest), $k=5$ and $k=8$ (richest) at the three locations. }
\label{fig:dynamics}
\end{figure}

We perform our analysis by averaging variables over time. Indeed, as already mentioned we are interested in long-term trends and segregation patterns, so it makes sense to ignore high-frequency fluctuations. Moreover, in this section we show that at least some variables quickly reach a steady state punctuated by noise. Other variables display larger fluctuations but these are mostly driven by noise. The lack of economically meaningful endogenous dynamics in our model further justifies taking time averages.

In Figure \ref{fig:dynamics}A we show the time evolution of the market price at three locations. At $X=(0,0)$ and $X=(2,2)$ the market price reaches a relatively stable value after about 25 time steps, after which the fluctuations are typically around 5 price units. The market price at $X=(4,4)$ reaches a similar state, but the transient is longer, of the order of 75 time steps. Unsurprisingly, the market price is highest in the center, where the attractiveness is maximal, and decreases moving farther from the center. 

The dynamics of the attractiveness is shown in panel B of the same figure. At initialization ($t=0$), the agents are allocated randomly across the city, irrespective of their income category. Therefore, up to noise the average income $\overline{Y}(X,0)$ at any location $X$ is equal to the average income over the city $\overline{Y}(0)$. This means that the social component of the attractiveness is (up to noise) $A^S(X,t)=1$, so at initialization the attractiveness $A(X,t)$ approximately corresponds to the intrinsic component, $A(X,0) \approx A^0(X)$. As the social composition of neighborhoods changes over time, so does the attractiveness. At $X=(0,0)$ it raises steadily from $A(X,0)=1$ to $A(X,t>50)\approx 1.4$, while at $X=(2,2)$ the attractiveness also rises but by a smaller amount. In both locations the increase of $A(X,t)$ is explained by the share of (relatively) rich agents increasing over time, creating a positive feedback that further increases the share of rich agents.   The opposite mechanism is at play at $X=(4,4)$, where the share of rich agents decreases over time. (Note that because $A^0(X=(4,4))$ is two orders of magnitude smaller than at $X=(0,0)$, for visualization purposes we show the attractiveness at $X=(4,4)$ on the right scale of the vertical axis.) In all cases, the social attractiveness quantitatively plays a minor role with respect to the intrinsic attractiveness.

A more detailed representation of the evolution of the social composition of neighborhoods is given in Figure \ref{fig:dynamics}C. We show the number of agents whose status is either housed or seller and belonging to three income categories, in the same three locations as above. At $X=(0,0)$ the poorest of the three categories (3-agents, in red) disappears after 50 time steps, because the market price becomes larger than their income $Y_3$. The number of agents belonging to the two other categories (5-agents, in cyan, and 8-agents, in green) instead fluctuates over time. These numbers also fluctuate at $X=(2,2)$ and $X=(4,4)$, with the share of 3-agents increasing as one moves farther from the center. 

\begin{figure}[!ht]
\centering
\includegraphics[width=0.5\textwidth]{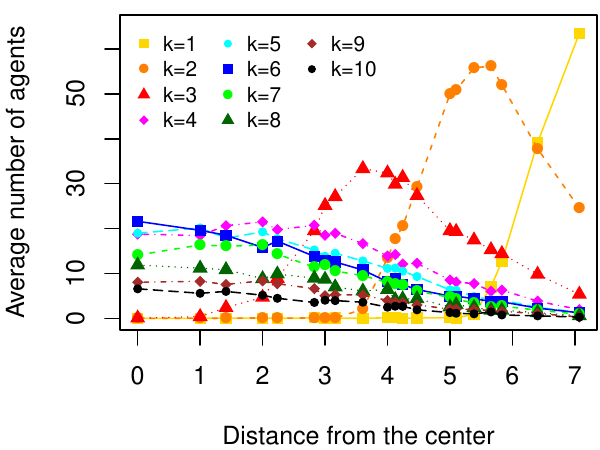}
\caption{\textbf{Spatial distribution of \textit{k}-agents as a function of the distance from the center.} The parameter values are the same as in Fig. \ref{fig:dynamics}. The number of \textit{k}-agents is averaged over 100 time steps over all locations at the same distance from the center.}
\label{fig:distribution-housed}
\end{figure}

Finally, in Figure \ref{fig:distribution-housed} we show the time-averaged shares of $k$-agents as a function of the distance from the center. (We consider a transient of 50 time steps, and then average from time step 51 to time step 150.) This figure shows that the categories with lowest income ($k=1$--$3$) are segregated out of the center, while agents from the other categories are mostly located up to distance $r=4$. However, the shares of $k$-buyers ($k>3$) at locations $r<4$ are not proportional to the population shares. For example, at $r=0$ the share of 10-agents is roughly 1/3 of the share of 4-agents, but in the population of buyers $\Gamma_{10}/\Gamma_4=1/5$ (see Appendix \ref{sec:parvalues}), so 10-agents are over-represented in the center. This is caused by the continuous double auction: although the market price is below the reservation price of 4-buyers, 10-buyers bid higher and are matched first. Because in the very center demand largely exceeds supply, 10-agents are more likely to secure an apartment.\footnote{The over-representation is also partly caused by a substitution effect: 4-buyers would spend most of their income to live at $r=0$, so they prefer neighborhoods farther from the center where they can afford a higher level of non-housing consumption.}

\subsection{Effect of the income distribution}
\label{sec:income}

We now study the effect of the income distribution of the buyers on the prices and segregation patterns. We consider twelve income distributions, with increasing levels of inequality. We keep the shares of $k$-buyers $\Gamma_k/\Gamma$ fixed, and vary instead the minimum income $Y_1$ and the total income spread $\Delta$. Indeed, buyers categories are arbitrarily defined, and it is the relative income spread $\Delta/Y_1$ that determines the level of inequality (see below). The important constraint is that the total income of the buyers must be the same across income distributions to allow for a meaningful comparison.

The total income of the buyers is 
\begin{equation}
M  = \sum_{k=1}^K Y_k \Gamma_k = Y_1 \Gamma + \Delta \sum_{k=2}^K (k-1)\Gamma_k = \Gamma \left[ Y_1 + \Delta \sum_{k=2}^K (k-1)\Gamma_k/\Gamma \right].
\label{eq:totincome}
\end{equation} 
In order to quantify income inequality, we take one of the possible definitions of the Gini index, namely half the relative absolute mean difference of incomes \citep{cowell2000measurement}. With the discrete distribution we consider, the Gini index reads
\begin{equation}
G   = \frac{\sum_{k=1}^K{\Gamma_k \sum_{j=1}^K \Delta |k-j| \Gamma_j }}{2\Gamma M} .
\label{eq:gini}
\end{equation} 
We keep $M=60\Gamma$ and fix the shares of $k$-buyers to $\{\Gamma_k/\Gamma\}=( 0.25,0.20,0.15,0.10,$ $0.08,0.07,0.06,0.04,0.03,0.02)$. We then vary $\Delta$ and $Y_1$ to obtain twelve different income distributions, whose Gini coefficients range from $G=0.26$ to $G=0.48$. For example, the most equal income distribution ($G=0.26$) is characterized by $Y_1=30$ and $\Delta=11.86$, so the relative income spread is only $\Delta/Y_1=0.4$ and the income of the richest buyers is larger by a factor of 4.5 than the income of the poorest. Conversely, the most unequal income distribution ($G=0.48$) has $Y_1=5$ and $\Delta=21.74$, so $\Delta/Y_1=4.35$ and the richest buyers have 40 times the income of the poorest. All values of $\Delta$, $Y_1$ and $G$ are specified in Appendix \ref{sec:parvalues}.

It is not meaningful to compare different income distributions by considering the spatial distribution of the shares of $k$-agents (as in Figure \ref{fig:distribution-housed}). Indeed, these categories are arbitrarily defined and we need to measure segregation in a way that is independent of the level of inequality. Therefore we resort to a \textit{rank-order information theory index} \citep{reardon2011income} which only uses information about the rank ordering of incomes, and is thus independent of the income distribution. We denote income percentile ranks by $p\in [0,1]$. For any given value of $p$, we calculate the segregation over the city between households with income ranks less than $p$ and households with income ranks greater or equal to $p$. We then average over all values of $p$. More specifically, denote by 
\begin{equation}
E(p)=-p\log_2 p - (1-p)\log_2 (1-p)
\end{equation}
the information entropy of the population when divided into these two groups, and by
\begin{equation}
H(p)=1-\frac{1}{L^2}\sum_{X \in \Omega} \frac{E_X(p)}{E(p)}
\end{equation}
the Theil index of segregation in the population divided between these two groups, where $E_X(p)$ is the information entropy calculated at location $X$. The rank-order information theory index $H^R$ is then
\begin{equation}
H^R=2\ln 2 \int_0^1 E(p)H(p)dp
\label{eq:Hr}
\end{equation}
This quantity varies between a minimum of zero, which corresponds to complete lack of segregation (the income distribution in each location $X$ mirrors the global income distribution, so that $E_X(p)=E(p), \forall X \in \Omega$), and a maximum of one with complete income segregation (in every location all households have the same income).

\begin{figure}
\centering
\includegraphics[width=1.\textwidth]{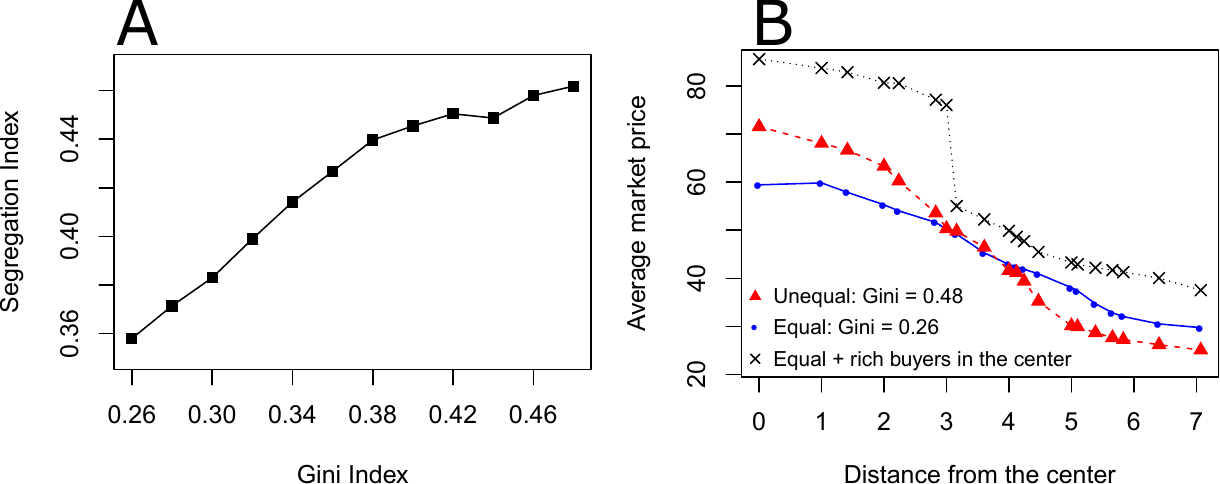}
\caption{\textbf{Effect of the income distribution on segregation and prices.}  (A) Rank-order information theory segregation index $H^R$ (Eq. \ref{eq:Hr}) as a function of the Gini Index (Eq. \ref{eq:gini}). Higher inequality leads to more segregation, in accordance with empirical evidence. (B) Average market price for the most equal and most unequal income distributions, and with an additional influx of rich buyers who search for an apartment only in the center. In the latter case, the prices increase all over the city and not only in the center. Globally, the prices are lower if the income distribution is unequal.}
\label{fig:sim-quantities}
\end{figure}

In Figure \ref{fig:sim-quantities}A we show how the segregation index $H^R$ varies with the level of inequality, as measured by the Gini index. We see that segregation increases with inequality, most strongly for $G<0.38$. There is little empirical evidence on the effect of income inequality on income segregation, because spatial income data are rarely available. Most studies \citep{wheeler2008trends,watson2009inequality,reardon2011income} rely on U.S. census data, because binned income distributions (with 15-25 bins) are available for each census tract. These studies find that income inequality increases income segregation, in accordance with our model. \cite{reardon2011income}  attribute this result to income-correlated social preferences, or to higher provision of local public goods in neighborhoods where the rich live \citep{tiebout1956pure}. Here we show that our ABM can replicate this finding.

We look at the spatial distribution of prices in Figure \ref{fig:sim-quantities}B. Comparing the most unequal and the most equal income distributions, in the unequal case the prices are higher in center and lower in the peripheries. Indeed, in the outskirts the share of lowest-income households is larger if the income distribution is unequal, reducing the prices. Because most agents reside in the locations where the prices go down, the global effect on prices is negative (-4\%).  This result is in line with \cite{maattanen2014income}, who come to the same conclusion using an assignment model that determines an \textit{equilibrium price gradient}. The authors also show that this finding is supported empirically in a number of U.S. cities. Using their model, they perform a counterfactual exercise and calculate what the prices would be if the level of inequality had not increased. They find values between 0 and 10\% higher according to the specific metropolitan area, in quantitative accordance with our model. 

We also experiment with an additional influx of rich agents in the center, testing whether the prices only grow in this location or whether they increase all over the city. We mimic the process by which rich households coming from outside -- the ``foreigners'' -- purchase real-estate properties in a city, either as a luxury good or as a secondary residence \citep{cvijanovic2018real}.\footnote{Another important category of wealthy external buyers is investors, who are mostly driven by interest and exchange rates and by home market conditions \citep{cvijanovic2018real}. As already mentioned, in this paper we focus on long-term outcomes. Because investor dynamics is highly volatile we do not think of agents composing the additional influx of buyers as investors.} These buyers usually choose the most attractive locations, and distort the local housing market because of their disproportionately high reservation prices \citep{chinco2015misinformed}. 

We assume that $\Gamma/10$ ``foreigners'' try to purchase a property at any time step $t>0$ with uniform probability in all locations within a radius $r=3$ from the origin $O$, i.e. $\forall X = (x,y) \in \mathds{Z}^2 \text{  s.t.  } x^2+y^2 \leq 9 $. We impose this hard threshold because we will assess whether prices and segregation patterns change all over the city and not just in the center. Therefore, we need to exclude the trivial scenario in which foreigners directly affect these variables by searching all over the city. The income of the foreigners is $Y_K + \Delta$, that is larger by a factor of $\Delta$ than the income of the $K$-agents. Other than that, the foreigners participate to the housing market following the same protocol as the other agents, e.g. they are not necessarily matched first in the order book.

We consider the most equal income distribution as a benchmark (but the results are robust to the choice of the distribution). Figure \ref{fig:sim-quantities}B shows that the prices increase substantially in the area where the foreigners search, but also all over the city, especially in the peripheries. The arrival of foreigners in the center leads to price growth in the most attractive locations. In turn, the increase in prices makes these locations less appealing to high-income households. Indeed, recall from Section \ref{sec:demand} that households also value non-housing consumption, and so are willing to substitute the attractiveness for cheaper locations where they can afford a higher consumption level. As the high-income households move to less attractive locations, their bids push up the prices there as well. So middle-income households may decide to search in the least attractive locations, and the process cascades all over the city in the steady state. This implies that 1-agents (the poorest buyers) cannot afford buying properties even in the peripheries and are segregated out of the city, differently from the benchmark case without additional influx in Figure \ref{fig:distribution-housed}. (This is not shown explicitly here but can be checked in the replication files.)

These conclusions are supported by indirect empirical evidence. For example, \cite{cvijanovic2018real} analyze the purchases of foreigners in the Paris housing market  and show that foreigners crowd out residents, overpay and cause prices to increase in the most attractive locations. \cite{sa2016effect} finds that foreign investment in England and Wales has a positive causal effect on house price growth at different percentiles of the distribution, but she does not consider the spatial aspect. From a theoretical point of view, \cite{favilukis2017out} come to our same conclusions. They use an overlapping generations model in which heterogeneous households decide consumption, savings, labor supply, tenure status, and location. In equilibrium, the households anticipate the arrival of the foreigners (out-of-town home buyers), and adjust their decisions accordingly.

\subsection{Effect of subsidies and taxes}
\label{sec:subtax}

We investigate which policy is most effective at reducing income segregation and increasing social welfare. Housing market policies have traditionally been divided in two strands. 

First, subsidized housing aims at improving the accessibility of low-income households to the housing market. There are two types of subsidies -- project-based and tenant or buyer-based \citep{sinai2005low}. The former include public housing and subsidies to the construction sector, which is incentivized to construct new affordable houses. The latter include buyer-based vouchers, certificates, rent supplements etc. There is a strong consensus in the literature for buyer-based subsidies \citep{olsen2003housing}, and so we focus on this policy (also, we do not consider constructions in our model).

Second, transaction taxes increase the cost of transacting a dwelling. To avoid paying the tax, many households may postpone buying or selling a dwelling \citep{dachis2011effects}, or adjust the transaction price to exploit discontinuities in tax liability \citep{best2017housing}. For our purposes, transaction taxes are also aimed at ``cooling'' the housing market and therefore improve affordability for low-income households, as confirmed by the recent 15\% transaction tax on foreign purchases in Vancouver.\footnote{See \texttt{https://www2.gov.bc.ca/gov/content/taxes/property-taxes/} 

\texttt{property-transfer-tax/understand/additional-property-transfer-tax}.}  Transaction taxes can be on buyers (in the form of stamp duty taxes) or on sellers (capital gains tax). We focus here on transaction taxes on buyers, which are widespread in OECD countries \citep{van2005new}.

We implement a system of \textit{ad-valorem} taxes and subsidies on buyers in a stylized fashion. We denote by $\xi_k$ the tax or subsidy for $k$-agents, where $\xi_k>0$ indicates a tax, and $\xi_k<0$ denotes a subsidy. For example, a 10\% subsidy corresponds to $\xi_k=-0.1$, and a 10\% tax to $\xi_k=0.1$. The budget constraint of the buyers becomes $z_k(t) + (1+\xi_k) P(X,t) = Y_k$. Replacing the budget constraint in the utility function, we get the indirect utility
\begin{equation}
V_k^\xi(X,t) = \left( Y_k - (1+\xi_k) P(X,t)  \right)^{1-\beta} \left(A(X,t) \right)^\beta.
\label{eq:utilitytaxes}
\end{equation}
The reservation demand prices are also affected by the tax or subsidy, with $P^d_k = Y_k / (1+\xi_k$). For instance, with a 10\% subsidy, the purchasing power of agents with income $Y=15$ rises to $Y/0.9=16.7$. On the contrary, with a 10\% tax, the purchasing power of agents with the same income reduces to $Y_1/1.1=13.6$.

We analyze the effect of three policies on the prices and segregation patterns, and compare with the no policy benchmark. The magnitude of subsidies and taxes should not be taken literally, as we are interested in qualitative differences and are not calibrating the model against real data. The policies are:
\begin{itemize}
\item Subsidies only: $\xi^S=(-0.20,-0.15,-0.10,-0.05,0.00,0.00,0.00,0.00,0.00,$ $0.00 )$. 1-agents receive a 20\% subsidy, 2-agents receive a 15\% subsidy, etc.
\item Taxes only: $\xi^T=(0.00,0.00,0.00,0.00,0.00,0.00,0.05,0.10,0.15,0.20)$. We implement an income-dependent transaction tax, with 10-agents paying 20\% of the transaction price, 9-agents paying 15\%, etc.
\item Subsidies and taxes: $\xi^{ST}=(-0.20,-0.15,-0.10,-0.05,0.00,0.00,0.05,0.10,$ $0.15,0.20)$. We combine the systems of taxes and subsidies. With this policy subsidies are funded through transaction taxes.
\end{itemize}

\begin{figure}
\centering
\includegraphics[width=1.\textwidth]{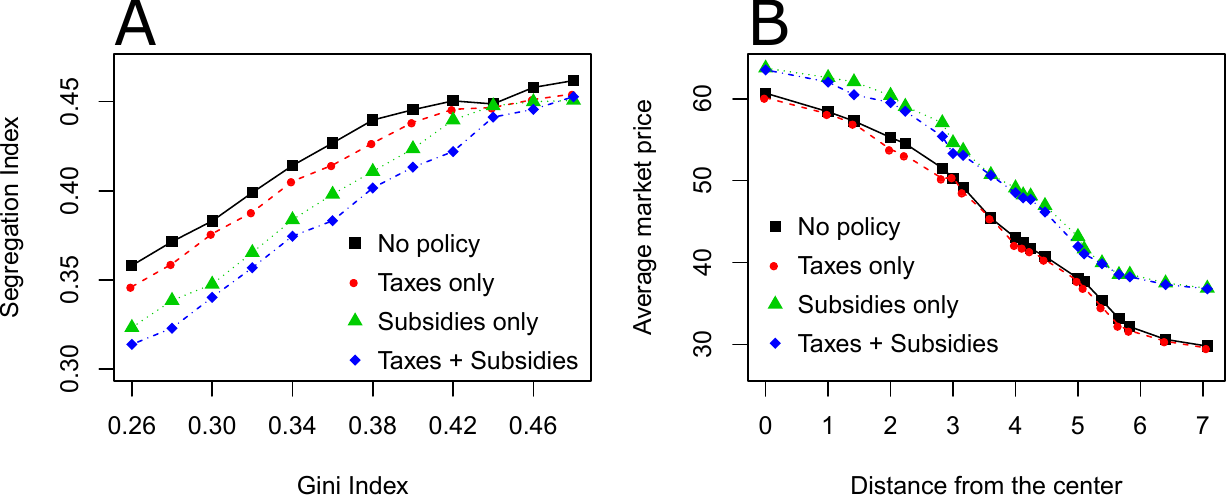}
\caption{\textbf{Effect of subsidies and taxes on segregation and prices}. (A) Segregation (as quantified by the rank-order information theory index $H^R$) as a function of the level of inequality for the various policies. The no policy benchmark corresponds to Figure \ref{fig:sim-quantities}A. (B) Average market price as a function of the distance from the center for various policies and in the case of the most equal income distribution (the no policy benchmark corresponds to the price shown in Figure \ref{fig:sim-quantities}B for $G=0.26$).}
\label{fig:segr-quantities}
\end{figure}

Figure \ref{fig:segr-quantities} illustrates the results of our policy exercise. In Figure \ref{fig:segr-quantities}A we see that taxes have a very small effect in mitigating income segregation, whereas subsidies are more effective (and if used in combination with taxes yield the lowest segregation). The effect of policies is more pronounced if the level of income inequality is not too high.

Figure \ref{fig:segr-quantities}B shows the average market price for the various policies. In this model, taxes are not successful at decreasing the prices. There are at least two reasons for this. First, the reservation prices of the richest buyers -- who are the ones mostly affected by the taxes -- are still above the market price at any location, so the richest buyers do not pay a fundamentally different price. Second, because the global demand $\Gamma$ is assumed inelastic, taxes cannot crowd out rich buyers. Figure \ref{fig:segr-quantities}B also shows that subsidies increase the market price. However, the rise is included between 4 and 6 price units, whereas the growth in purchasing power that the subsidies entail for the lowest-income households is 7.5 price units ($Y_1/\xi_1=30/0.8=37.5$). So subsidies make it possible for these households to afford properties in locations which they could not afford without the policy.

\begin{figure}
\centering
\includegraphics[width=0.5\textwidth]{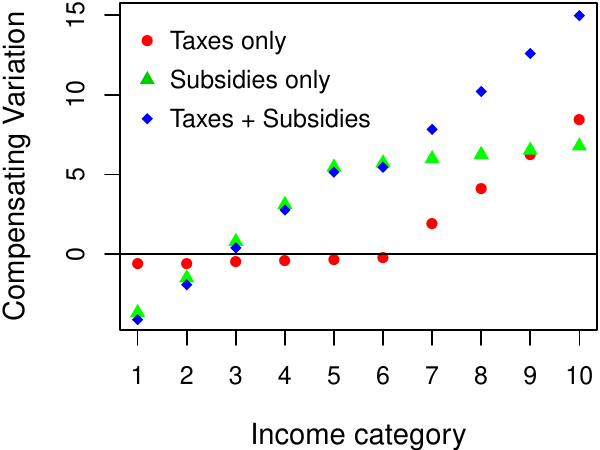}
\caption{\textbf{Welfare effects of subsidies and taxes.} The Compensating Variation (CV) is the amount of money that an agent should receive to maintain the same utility level after the introduction of the policy. Therefore, a positive CV is associated with a decrease in utility, while the opposite is true when the CV is negative. The CV subsumes the different channels through which the policies impact the welfare of the agents (see Eq. \eqref{eq:compvar}), but it does not consider the long-term benefits of social mixing. All policies increase the welfare of the agents with lowest income categories, at the cost of decreasing the welfare of richer agents. }
\label{fig:welfare}
\end{figure}

So far we have only compared the different policies based on the effect they have on the segregation index $H^R$ and on the market prices. We conclude this section by performing a welfare analysis of these policies. By changing the prices and the social composition of neighborhoods, these policies impact the utility functions of all agents. The \textit{compensating variation} $CV_k^\xi(X)$ for $k$-agents at location $X$ due to the introduction of policy $\xi$ is the amount of money that the same agents would need to receive to return to the original utility level \citep{mas1995microeconomic}. Denoting $V_k^0\left(Y_k,P(X),A(X)\right)$ as the indirect utility for $k$-agents at $X$ without the introduction of the policy, the compensating variation is defined by
\begin{equation}
V_k^0\left(Y_k,P(X),A(X)\right)=V_k^\xi\left( Y_k^\xi + CV_k^\xi(X),P^\xi(X),A^\xi(X) \right),
\label{eq:compvarcond}
\end{equation}
where $V_k^\xi$, $Y_k^\xi$, $P^\xi(X)$ and $A^\xi(X)$ denote the same variables with the introduction of the policy. Note that the prices, the attractiveness and also potentially the compensating variation should also depend on time. As we take time averages, to simplify the notation we drop the time dependence.

Solving Eq. \eqref{eq:compvarcond} for $CV_k^\xi(X)$ using the usual form of the utility function and the usual specification of subsidies and taxes (see Eq. \eqref{eq:utilitytaxes}) yields
\begin{equation}
CV_k^\xi(X)=\xi_k P^\xi(X) + \left(P^\xi(X)-P(X)\right) + \left(Y_k - P(X)\right)\left[\left(\frac{\overline{Y}(X)}{\overline{Y}^\xi(X)}\right)^{\frac{\beta}{1-\beta}}-1 \right].
\label{eq:compvar}
\end{equation}

The three terms on the right hand side of the expression above clarify the channels through which the different policies impact the welfare of the agents. The first term is the amount of money necessary to compensate the increase (decrease) in purchasing power that the subsidy (tax) entails. Indeed, when $\xi_k>0$ (tax), this term is positive, while when $\xi_k<0$ this term is negative (the agent's utility increases, so the agent should give money and not receive money to keep the same level of utility). The second term is the difference in prices. When the prices increase, due to the introduction of the policy, $P^\xi(X)-P(X)$ is positive and agents need to receive money to keep the same level of utility. If the prices decrease because of the policy the opposite if true. Finally, the third term is positive if $\overline{Y}(X)/\overline{Y}^\xi(X)>1$, i.e. if the policy reduces the average income at location $X$. This implies a reduction in utility through the social component. An increase in average income would instead increase utility and thus lead to a negative contribution in the compensating variation.

One thing that $CV_k^\xi(X)$ does not capture is the long-term benefits of social mixing \citep{benabou1993workings}. Indeed, we do not have a labor market or an educational system in our model, and the social composition of neighborhoods only enters utility through the average income. Therefore, we should expect that the compensating variation is overestimated in our analysis, i.e. the decrease in welfare due to the various policies does not take into account the long-term increase in welfare due to a better functioning labor market and educational system.

In Figure \ref{fig:welfare} we show the compensating variation $CV_k^\xi =\sum_X CV_k^\xi(X) / L^2$, averaged over all locations $X$, for $k$-agents for policy $\xi$. In all cases this quantity is negative for agents belonging to the lowest income categories and positive for all other agents, suggesting that only the welfare of the poorest agents improves. A utilitarian social welfare function assumes that social welfare is just the sum of the welfare of all agents: $SW^\xi_U=\sum_k \Gamma_k CV_k^\xi / \Gamma$. From a utilitarian point of view, the global welfare would decrease due to the introduction of the various policies. According to a Rawlsian welfare function $SW^\xi_R=\min_k CV_k^\xi$ instead the global welfare improves, because the poorest agents are increasing their welfare. 

\section{Conclusion}
\label{sec:conclusion}

In this paper we have introduced a baseline Agent-Based Model (ABM) of the housing market. Our goal was to better understand the relation between segregation and income inequality, house prices and policies by explicitly representing the behavior of buyers and sellers and the market dynamics. In our model, the behavioral rules rely on discrete choice theory and on the fast and frugal heuristics paradigm, and the price formation mechanism is represented using a continuous double auction. We need not impose any equilibrium restriction (such as finding the conditions for which all agents have the same utility, as in spatial equilibrium models) because in the ABM methodology outcomes are the result of unconstrained interactions among agents. 

We have found a partial analytical solution of the ABM. From a methodological point of view, this shows that ABMs are not incompatible with mathematical analyses, and that equilibrium and rationality are not simplifying assumptions necessary for mathematical tractability. Our analytical solution makes it possible to analyze the structure of the model and the effect of the parameters. For instance, a closed-form solution shows that increasing the number of apartments available in a city reduces income segregation.

We have simulated the model to study the interplay between the income distribution of the buyers and the spatial price distribution, with a focus on the segregation patterns. We have then analyzed the global effect of a demand spike localized in the center, and we have compared a number of policies whose goal is to foster social mixing. Some of our results are in line with other findings in the literature, but our modeling methodology allows for a simpler narrative that directly matches the housing market dynamics. Other results account for previously unexplained empirical facts. For example, we are not aware of any theoretical contribution that reproduces the positive causal link from income inequality to income segregation. 

Our model can be extended in several ways, depending on the research question. Because of the enhanced flexibility implied by the lack of equilibrium constraints, it is extremely easy to add more realistic features (at least in the numerical simulations).  For example, we may endogenize the global demand and the leaving rate. Coupled with a role for expectations, this is likely to generate price cycles in the city. We can also endogenize the intrinsic attractiveness, e.g. to consider agglomeration effects. We can finally model mortgages and a financial sector, so to investigate whether there is any implication of segregation on financial stability.

More importantly, this paper has served as a \textit{qualitative} demonstration that a disaggregated agent-based representation of the housing market can account for a wide range of both well-known and less obvious phenomena related to income segregation. The next step to show that ABMs may be a more faithful representation of real housing markets (than e.g. spatial equilibrium or assignment models) would be to show that ABMs better predict prices and segregation patterns \textit{quantitatively} and out of sample. Some works \citep{filatova2015empirical,baptista2016macroprudential} are starting to bring housing ABMs to data, but much more work is necessary in this direction.

\appendix

\section{ODD+D description of the Agent-Based Model}
\label{sec:oddd}

Having a standard for describing Agent-Based Models (ABMs) makes comparisons across ABMs easier and helps researchers describe their models more clearly. To this end, \cite{grimm2010odd} proposed a protocol named Overview, Design Concepts and Details (ODD), which was supplemented by \cite{muller2013describing} to include human Decision making (hence the protocol got named ODD+D). In this appendix we describe our ABM following the ODD+D protocol.

\subsection{Overview}

\subsubsection{Purpose}

In this paper we introduce an ABM of the housing market. The purpose of our ABM is to understand phenomena related to income inequality and income segregation and to study the effects of policies designed to tackle these. An ABM is well-suited to this goal because it makes it easy to fully consider heterogeneity by modeling the market in a decentralized way without imposing aggregate constraints such as equilibrium. 

\subsubsection{Entities, state variables and scales}

The agents in our model are households. They are mainly characterized by their \textit{state}: they can be buyers, sellers, or housed if they are neither buying nor selling. Agents are also characterized by their income. The exogenous driving factor of our model is the structure of the city, here represented by an intrinsic attractiveness that subsumes amenities and the convenience of the transportation system. This exogenous factor is supplemented by an endogenous social component that depends on the social composition of neighborhoods. Space is a grid of fixed size with a finite number of locations. Time is discrete and the time horizon is infinite.

\subsubsection{Process overview and scheduling}

At the beginning of each time step buyers come to the city from the outside and select a location in which they will search for a dwelling. At the same time, housed agents may decide to put their dwelling on sale with a certain probability, and join the set of sellers at the same location. Next, at each location transactions take place mediated by a continuous double auction. Finally, successful buyers take residence in the location where they searched, successful sellers leave the city and the updated market price is computed.

\subsection{Design Concepts}

\subsubsection{Theoretical and Empirical Background}

Our ABM builds on a substantial body of work on ABMs of the housing market, and most of our assumptions are drawn from the existing literature (see the literature review in the Introduction). With respect to many ABMs in this literature, we designed our model to be parsimonious and tractable. The choice of the behavioral rules is based on the fast and frugal heuristics paradigm \citep{gigerenzer1999simple}. This suggests that in environments in which information is limited and dispersed using simple heuristics may be optimal.

\subsubsection{Individual Decision-Making}

Buyers select a location where they could search for a dwelling based on the expected utility at that location. However, instead of maximizing their expected utility, they choose a location with a probability proportional to the utility, in the same spirit as in discrete choice theory. Sellers employ an aspiration level heuristic, by which they try to apply a markup on the current market price and reduce their requested price successively if they fail to sell their dwelling. 

\subsubsection{Learning}

No learning takes place.

\subsubsection{Individual Sensing}

Buyers are expected to know the market price and the attractiveness in each location, further to their own income. The fact that they choose a location with a probability proportional to the expected utility at that location may include imperfect measurement of the characteristics of a location -- buyers may fail to choose the optimal location because they may have limited information. Sellers only need to know the market price of the location where they live at the time in which they put their dwelling on sale.

\subsubsection{Individual Prediction}

The agents do not predict future conditions. Expectations would be particularly important in case of speculation, but this is not our focus here.

\subsubsection{Interaction}

Interactions between buyers and sellers are mediated by a continuous double auction. This is meant to represent bilateral bargaining between two groups in a stylized fashion.

\subsubsection{Collectives}

There are no collectives in this model.

\subsubsection{Heterogeneity}

Buyers are heterogeneous for what concerns their income. This impacts their decision making because poor households will not attempt at purchasing properties in expensive neighborhoods. Sellers are heterogeneous for what concerns their reservation price -- sellers whose dwelling has been on the market for a long period will accept a lower reservation price. 

\subsubsection{Stochasticity}

Initialization is completely random, but the model is ergodic so initialization does not matter. Buyers search among locations in a stochastic manner.

\subsubsection{Observation}

We let the model run for an initial transient and then take measurements when it has reached the steady state. We mostly measure prices and the number of agents who fall in distinct income categories in each location.

\subsection{Details}

\subsubsection{Implementation Details}

The model has been implemented in \texttt{NetLogo} \citep{wilensky1999netlogo}. The \texttt{NetLogo} code and the \texttt{R} code that runs the model through the RNetLogo interface \citep{thiele2012rnetlogo} are available on \texttt{Zenodo}: \texttt{https://dx.doi.org/10.5281/zenodo.1453347}.

\subsubsection{Initialisation}

All locations are initialized with a price that is below the minimum reservation price of the buyers. Buyers with different incomes are allocated randomly at initialization. 

\subsubsection{Input Data}

The model does not take data from external sources.

\subsubsection{Submodels}

For more details, we refer the reader to the code, and to Appendix \ref{sec:parvalues} for the parameter values and rationale for choosing those values.

\section{Mathematical derivations}
\label{sec:mathderivations}

\subsection{Derivation of the market price in the one category case}
\label{sec:mathderivationsonecat}

We proceed in three steps. First, we calculate the densities of buyers and sellers at any distance $r$, in the steady state. Second, we compute the average reservation demand and offer prices. Third, we obtain the market price.

We start calculating the expected steady state densities of buyers and sellers at distance $r$. We denote them by $n_b^\star(r)$ and $n_s^\star(r)$ respectively. The density of buyers is obtained multiplying the number of buyers $\Gamma$ by the probability density to choose a location at $r$ (Eq. \ref{eq:choiceprob}), given by the attractiveness $A(r)$ and by a normalization factor $Z$:

\begin{equation}
n_b^\star(r)=\Gamma A(r)/Z,\text{ with } Z=2\pi \int_0^{R_{max}}{r A(r) dr}.
\label{eq:nbss}
\end{equation}
The density of sellers $n_s(r,t)$ can be computed by summing the density of apartments already on sale, denoted by $\bar{n}_s(r,t)$, and the expected fraction of apartments newly put on sale: $n_s(r,t)=\bar{n}_s(r,t)+\alpha \left( n - \bar{n}_s(r,t) \right)$. By definition the apartments already on sale at time $t$ are those that were not sold at time $t-1$.
 
In order to calculate $\bar{n}_s(r,t)$ we make the crucial assumption in this mathematical derivation, namely that all buyers at location $r$ and time $t-1$ succeed in securing an apartment. This assumption is correct only if two conditions are met. First, the number of buyers must be smaller or equal than the number of sellers. Second, all buyers must afford the dwellings, that is the reservation demand price of all buyers must be larger than the reservation offer price of all sellers. The validity of this assumption is checked \textit{ex-post} in the discussion in Section \ref{sec:M1}.

 In any case, if all buyers secure an apartment, we have $\bar{n}_s(r,t)=n_s(r,t-1)-n_b(r,t-1)$. In the steady state, 
\begin{equation}
n_s^\star(r) = n - \frac{1-\alpha}{\alpha}  n_b^\star(r).
\label{eq:nsss}
\end{equation} 
We now calculate the expected reservation prices. Since there is only one income category, the reservation demand prices are all identical, and correspond to $Y=Y_1$. On the contrary, the reservation offer prices are heterogeneous, because they depend on the time on market (Eq. \ref{eq:alh}). Here we make a slightly simplifying assumption, namely that the sale probability is the same for all dwellings on the market at location $r$, and therefore corresponds to the \textit{market tightness} $q^\star=n_b^\star/n_s^\star$. In fact, cheaper dwellings are more likely to sell in the order book, but comparing to simulations in Section \ref{sec:M1} we show that this is a second-order effect. 

Using Eq. \eqref{eq:alh}, the expected steady state reservation offer price at a location at distance $r$ is 
\begin{equation}
\mathbb{E} \{ P^{s,\star} \} (r) = (1+\mu) P^\star(r) \mathbb{E} \{ \lambda ^{\frac{t-t_i}{\tau}} \}.
\label{eq:resofferprice}
\end{equation}
In the steady state, at each time step a dwelling on the market is sold with probability $q^\star$, and not sold with probability $1-q^\star$. We can calculate the expected discount $\mathbb{E} \{ \lambda^{k/\tau} \}$, with $k = t-t_i$, by using the geometric distribution:\footnote{Strictly speaking, the sellers decrease their reservation price at intervals of $\tau$ time steps, which is indeed what we do in the numerical simulations. However, in the analytical solution we are implicitly assuming that they decrease the reservation price by a factor $1/\tau$ every time step.}
\begin{equation}
\mathbb{E} \{ \lambda^{k/\tau} \} = \sum_{k=1}^\infty \left(\lambda^{1/\tau}\right)^{k-1} \left(1-q^\star\right)^{k-1} q^\star = \frac{q^\star}{1-\lambda^{1/\tau}\left(1-q^\star\right)}.
\label{eq:calcgeom}
\end{equation}
We can finally write the market price at distance $r$ from the center, in the steady state. We use Eqs. \eqref{eq:resofferprice} and \eqref{eq:calcgeom} and the definition of market price, that is a weighted average of the expected reservation and offer prices, with weight $\nu \in [0,1]$: 
\begin{equation}
P^\star(r) = \nu Y + (1-\nu) \mathbb{E} \{ P^{s,\star} \} (r).
\label{eq:lawofmotion}
\end{equation}
Rearranging gives Eq. \eqref{eq:mathfixedpricealpha}.

\subsection{Details on the one category case with general utility}
\label{sec:genutility}

The fixed point $P^\star(r)$ has the same functional form as in Eq. \eqref{eq:mathfixedpricealpha}, except that the ratio $A(r)/Z$ is replaced by $V^\star(r)/Z^\star$, where
\begin{eqnarray}\label{eq:utility1}
V^\star(r) = \left(Y-P^\star(r)\right)^{1-\beta} \left(A(r) \right)^\beta, \\
\label{eq:utility2}
 Z^\star = 2\pi \int_0^{R_{max}}{w \left( Y - P^\star(w) \right)^{1-\beta} \left(A(w) \right)^\beta dw}.
\end{eqnarray}
The problem is that we cannot explicitly solve for $P^\star(r)$ anymore, so we must use an iterative method. For each value of $\beta$, we start from the value of $Z^\star$ that we observe in the simulations, and solve Eq. \eqref{eq:mathfixedpricealpha}  (up to the transformation $A(r)/Z \rightarrow V^\star(r)/Z^\star$) numerically for 10000 values of $r$, $0<r<R_{max}$. We then numerically compute $Z^\star$ from Eq. \eqref{eq:utility2} with the trapezoidal method, and iterate this procedure until convergence for $Z^\star$ is reached. 

\section{Parameter values}
\label{sec:parvalues}

The baseline parameter values are reported in Table \ref{tab:ModelParameters}.

\begin{table*}[htbp]
	\centering
	\resizebox{0.7\textwidth}{!}{
		\begin{tabular}{|c|c|l|}
			\hline
			Symbol & Value & Description \\
			\hline
$N$ & 100 & Number of apartments at each location \\
$L$ & 11 & Linear size of the grid $\Omega$ \\
$a$ & 1 & Distance between neighboring locations \\
$R$ & 3 & Steepness parameter in the attractiveness \\
$K$ & 10 & Number of income categories \\
$Y_1$ & 15 &  Income of the lowest income category \\
$\Delta$ & 5 & Difference in income between two consecutive categories \\
$\beta$ & 0.5 & Weight given to the attractiveness in the utility function \\
$\alpha$ & 0.1 & Probability for housed agents to become sellers\\
$\mu$ & 0.1 & Markup that the sellers try to apply to the market price \\
$\lambda$ & 0.95 & Discount rate on the reservation offer price \\
$\tau$ & 2 & Time steps of unsuccessful sale before price revision \\
$\nu$ & 0.1 & Bargaining power of the seller \\
$\Gamma$ & 1000 & Total number of incoming agents each time step\\
			\hline
		\end{tabular}}
	\caption{Model parameters.}
	\label{tab:ModelParameters}
\end{table*}

We discuss the calibration choices below:
\begin{itemize}
\item $\lambda=0.95$. \cite{merlo2004bargaining} analyze one of the most detailed datasets on listing price changes and offers made between initial
listing and sale agreement. They find that, on average, the first price reduction is 5.3\% and the second reduction is 4.4\%. \cite{baptista2016macroprudential} also use $\lambda=0.95$, obtained from Zoopla data (Zoopla is a popular online portal for real-estate services in England). Finally, \cite{loberto2018potential} also find a similar value from the analysis of a housing advertisements website in Italy.
\item $\Gamma=1000$, $N=100$, $\alpha=0.1$. As shown in Eq. \eqref{eq:mathfixedpricealpha}, the price is determined (among other things) by the relative magnitude of these parameters. We chose these specific values to have a reasonable level of noise. Indeed, with too few agents (e.g. $\Gamma=100$, $N=10$) the random arrival of buyers would generate wild price fluctuations, and with too many agents (e.g. $\Gamma=10000$, $N=1000$) the price dynamics would almost be deterministic. Finally, note that in Section \ref{sec:math} we have used $\Gamma=400$ for illustrative purposes. 
\item $\beta=0.5$. The households value housing and non-housing consumption equally.
\item $L=11$, $R_{max}=6.21$, $R=3$, $a=1$. The attractiveness decreases at $R$ to approximately 1/3 of its value in the center.  With this choice, $R$ is close to half the radius of the city. Therefore, we have an attractive center and a non-attractive periphery, which is a necessary condition for price differentiation. $L$ (or equivalently $R_{max}$) and $a$ just determine the size of the city.
\item $\mu=0.1$, $\nu=0.1$. This parametrization captures the idea that the sellers have to post a price, which cannot be much higher than the market price. Therefore, most bargaining power is on the buyers' side.
\item $Y_1=15$, $\Delta=5$ and $K=10$. These parameters are chosen together to model a specific income distribution. The absolute magnitudes of $Y_1$ and $\Delta$ do not matter, only their ratio $\Delta/Y_1$ determines the inequality in the income distribution. While these values for $Y_1$ and $\Delta$ have been used to produce Figure \ref{fig:segregation-analytical}, in Table \ref{tab:paramincome} we list all the values that correspond to the twelve income distributions used in Section \ref{sec:num}.
\item $\{\Gamma_k/\Gamma\}=( 0.25,0.20,0.15,0.10,$ $0.08,0.07,0.06,0.04,0.03,0.02)$ is the income category distribution of $k$-buyers. When varying the income distribution we only change $\Delta$ and $Y_1$, keeping $\{\Gamma_k/\Gamma\}$ fixed.
\end{itemize}

\begin{table*}[htbp]
	\centering
	\resizebox{0.6\textwidth}{!}{
		\begin{tabular}{|ccc|}
			\hline
			Minimum income $Y_1$ & Income Spread $\Delta$ & Gini index \\
			\hline
30 & 11.86 & 0.26 \\
 			\hline
28 & 12.65 & 0.28 \\
 			\hline
26 & 13.44  & 0.30 \\
 			\hline
23.5 & 14.43  & 0.32 \\
 			\hline
21 & 15.41  & 0.34 \\
 			\hline
19 & 16.21  & 0.36 \\
 			\hline
16.5 & 17.19  & 0.38  \\
 			\hline
14 & 18.18  & 0.40 \\
 			\hline
12 & 18.97 & 0.42 \\
 			\hline
10 &  19.76 & 0.44 \\
 			\hline
7.5 & 20.75 & 0.46 \\
 			\hline
5 & 21.74 & 0.48 \\
		    \hline
		\end{tabular}}
	\caption{Parameters defining the income distribution.}
	\label{tab:paramincome}
\end{table*}

\bibliographystyle{C:/Users/marco/Dropbox/Latex/econ}
\bibliography{ref}

\end{document}